%% file: 202503_3DSAEF.tex
\documentclass[12pt,letterpaper]{article}
\input{packages}
\input{definitions}
\title{A three-dimensional energy flux acoustic propagation model}
\author[1]{Mark Langhirt}
\author[2]{Charles Holland}
\author[3]{Ying-Tsong Lin}
\affil[1]{Graduate Program in Acoustics, The Pennsylvania State University, 201 Old Main, University Park, PA 16802, USA, mal83@psu.edu}
\affil[2]{Department of Electrical and Computer Engineering, Portland State University, 1825 SW Broadway, Portland, OR 97201, USA, charles.holland@pdx.edu}
\affil[3]{Scripps Institution of Oceanography, University of California, 9500 Gilman Drive, La Jolla, CA 92093, USA, ytlin@ucsd.edu}
\date{}
\begin{document}
\maketitle
\begin{abstract}
  This paper extends energy flux methods to handle three-dimensional ocean acoustic environments, the implemented solution captures horizontally refracted incoherent acoustic intensity, and its required computational effort is predominantly independent of range and frequency.  Energy flux models are principally derived as incoherent solutions for acoustic propagation in bounded waveguides.  The angular distribution of incoherent acoustic intensity may be derived from Wentzel-Kramers-Brillouin modes transformed to the continuous angular domain via the ray-mode analogy.  The adiabatic approximation maps angular distributions of acoustic intensity as waveguide properties vary along a range-dependent environment, and the final solution integrates a modal intensity kernel over propagation angles.  Additional integration kernels can be derived that modulate the incoherent field by specific physical wave phenomena such as geometric spreading, refractive focusing, and boundary attenuation and interference.  This three-dimensional energy flux model is derived from a double-mode-sum cross-product, is integrated over solid-angles, incorporates a bi-variate convergence factor, accounts for acoustic energy escaping the computational domain through transparent transverse boundaries, and accumulates bottom attenuation along transverse cycle trajectories.  Transmission loss fields compare favorably with analytic, ray tracing, and parabolic equation solutions for the canonical ASA wedge problem, and three-dimensional adiabatic ray trajectories for the ideal wedge are demonstrated.
\end{abstract}

\section{INTRODUCTION}

The primary objective of this research was to extend the energy flux methods analytically to general three-dimensional (3D) underwater acoustic propagation problems that exhibit horizontal wave refraction.  The energy flux acoustic propagation model was first derived by Weston in 1959 \supercite{weston1959guided}, and independently derived from a different approach by Brekhovskikh in 1965\supercite{brekhovskikh1965average}.  Weston’s ``characteristic time'' quantity has come to be more commonly referred to as the ``ray invariant,'' and Milder demonstrated in 1969\supercite{milder1969ray} that it is an invariant of the ray system’s Lagrangian under the adiabatic approximation.  Weston elucidated on the energy flux model and methods in a series of papers in 1980\supercite{weston1980acoustica,weston1980acousticb,weston1980wave}, and both Weston and Tindle had been investigating the ray-mode analogy around the same time\supercite{tindle1979equivalence,weston1979reflection,tindle1980connection,tindle1980cycle}.  Weston and Harrison have both investigated potential applications of the ray invariant and energy flux methods in three-dimensional (3D) ocean acoustic propagation problems to a limited extent\supercite{weston1959guided,weston1961horizontal,harrison1977three,harrison1979acoustic}, but a fully functional 3D energy flux model based purely on energy flux methods has never been derived, implemented, and validated as of yet.

Harrison published in 2013\supercite{harrison2013ray,harrison2015efficient} an extension to the energy flux propagation model that reincorporated some of the inter-modal interference from neighboring modes.  A convergence factor integration kernel was introduced to the energy flux solution in order to capture refractive field structure on the scale of the cycle distances.  The convergence factor acts like an eigen-ray filter that focuses directionally dependent acoustic intensity at a particular receiver location when the adiabatic ray cycles converge.  This was the primary inspiration and motivation for the 3D solid-angle energy flux model presented in this paper, and the principal question investigated was, "can the refractive features captured by the convergence factor be applied to horizontal refraction in 3D environments?"

There are both benefits and drawbacks to using an energy flux model for computing three-dimensional underwater acoustic propagation, and they are plainly derived from the assumptions and approximations that are used to construct a tractable solution to an otherwise general partial differential equation.  Understanding the assumptions and approximations sets the appropriate context for a model's applicability when applying it to practical problems.  Since this 3D solid-angle energy flux model was derived purely from energy flux methods, all of the typical energy flux advantages and limitations also apply to this model.

Energy flux methods produce inherently incoherent solutions so they are generally inapplicable when coherence of the wave's phase is needed.  For applications where the average acoustic intensity is all that is required or reliable, then energy flux methods have the benefit of avoiding computations of fine-scale interference structure.  This effectively reduces the computational load of the model by avoiding the evaluations of summation cross-products of oscillatory functions.  The fact that fine-scale features of the acoustic field are ignored can very often be justified by considering that there is seldom sufficient ocean environmental information (even in 2D problems, let alone 3D) to support accurate fine-scale modeling.  Three-dimensional acoustic propagation problems are already very computationally demanding, so there is benefit in a solution that is more computationally efficient while still resolving the primary effects of horizontal refraction on the overall acoustic field.

It is also important to note that energy flux models do not require the use of marching methods, eigenvalue decomposition, or root-finding algorithms for solving the differential equations.  Furthermore, field calculations at individual receiver positions are independent of each other, and thus a direct source-to-receiver calculation is possible without computing the field at any other location.  The principal computational mechanism in energy flux models is a numerical quadrature routine across a continuous and real angular domain, and the reason for its considerable efficiency is that the integrands are slowly varying at all frequencies since the fine-scale interference structure is inherently neglected.  This may be contrasted with wavenumber integration techniques where the interference structure embedded within the integrand is highly oscillatory, which is a significant limitation governing the rate of a solution's numerical convergence as a function of the discretization resolution.

Section II outlines the derivation of the 3D solid-angle energy flux model and is divided into subsections that follow the analytical derivation and theory in a linear and consequential fashion.  Section III briefly outlines the implementation approach and contains comparisons for the canonical ASA wedge problem between this 3D solid-angle energy flux model, analytic solutions, parabolic equation solutions, and ray tracing solutions.  Section III also provides a demonstration of 3D adiabatic ray cycles computed from invertible mappings generated by partial cycle distance integrals.  Section IV discusses these energy flux developments from a theoretical perspective, the capabilities and limitations of this model, and potential avenues for future work. Lastly, section V summarizes the primary contributions and results from this work.

\section{MODEL DERIVATION}

Preliminary work by Langhirt\supercite{langhirt2025foundations} investigated potential theoretical and analytical mechanisms to lay the foundations for constructing a three-dimensional energy flux model, and that work lead directly to the model described in this paper.  Developments from that prior work included: interpreting the terms inside of the convergence factor as identifiable partial adiabatic ray cycles associated with particular source departure and receiver arrival directions, splitting the energy flux integration domain and associating corresponding convergence factor source cycle terms to enable asymmetric energy flux integration methods, the addition of a mode stripping kernel based on adiabatic ray cycles radiating from the source location to simulate free propagation through transparent boundaries, and simulating ray trajectories by inverting the mappings generated by partial cycle distance integrals.  All of these developments are necessary for the model described in this paper and are incorporated into its derivation.  This 3D solid-angle energy flux model is derived from a double (vertical and transverse) mode summation cross-product that is transformed to an integration over solid-angles.  Interaction with the seafloor is handled by continuous attenuation along the horizontal arc length of the transverse cycles.  The transverse boundaries of the computational domain are effectively transparent and thus a mode-stripping kernel nullifies acoustic intensity that arrives at these boundaries along the transverse adiabatic ray cycles.  The model also includes a bi-variate convergence factor that is derived using basic considerations of the model's differential topology.  

\subsection{Environmental description and derivation overview}

\begin{figure}[!htb]
  \includegraphics[width=\textwidth]{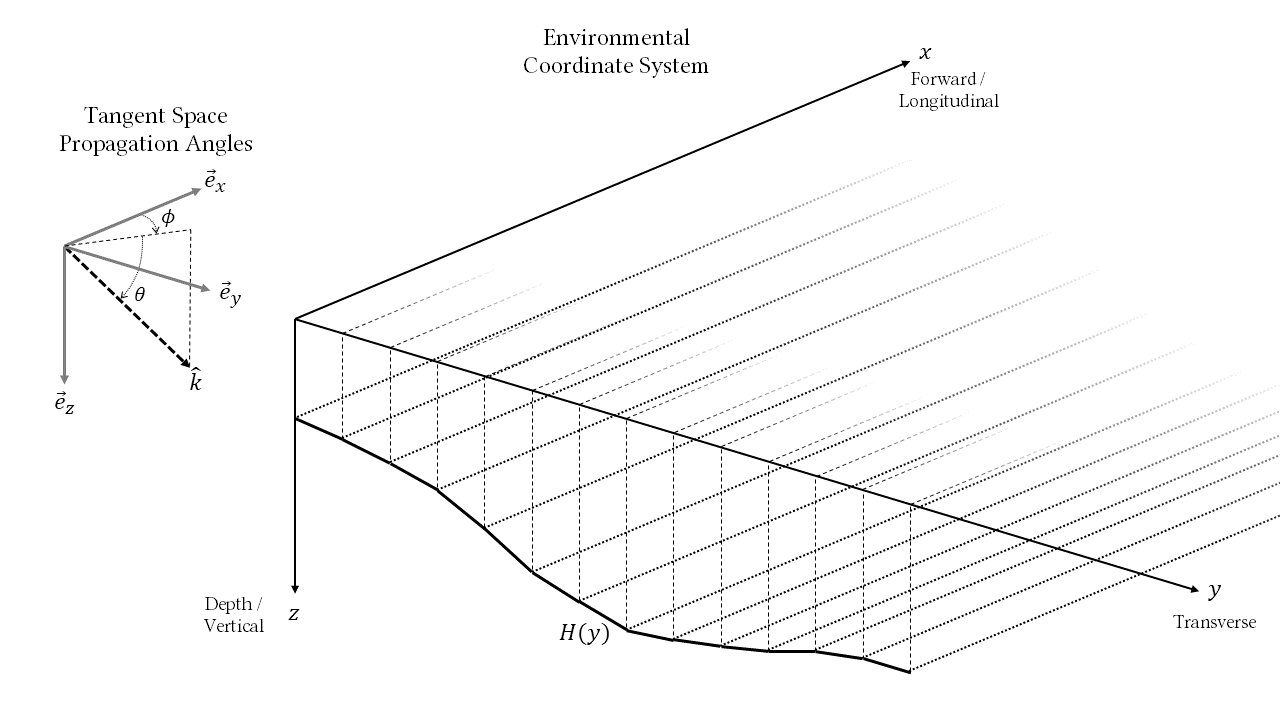}
  \caption{Right-handed Cartesian coordinates describing the environmental geometry, transverse ($y$-axis) range-dependence, and propagation angles for the 3D solid-angle energy flux model}
  \label{fig:EnvironmentCoords}
\end{figure}

Derivation of the 3D solid-angle energy flux model is similar to Harrison’s derivation of the 2D semi-convergent energy flux model\supercite{harrison2013ray,harrison2015efficient}.  The underlying coordinate system is chosen to be right-handed Cartesian with the ${}^+z$-direction pointing down from the sea surface, as shown in \cref{fig:EnvironmentCoords}.

The environment is range-dependent transversely along the $y$-axis, but is assumed to be translationally symmetric along the $x$-axis, which corresponds to the “forward” direction along which the model’s ray cycles are parameterized.  This constraint implies no back-propagation such that waves travelling forward in the ${}^+x$-direction will not turn around and their forward wavenumber components ($k_x$) are constant and strictly positive.  These environmental constraints allow enough generality to describe propagation along a wedge or trough while directly capturing the effects of horizontal refraction.

All  wavenumber profiles are assumed to be convex so that wave tunneling or beam bifurcations need not be considered, and the angular distribution of intensity mapped by the adiabatic transformation remains unimodal.  It is possible that convex vertical wavenumber profiles and convex transverse profiles for bathymetry and wavenumber could in some cases together produce non-convex transverse effective wavenumber profiles, but it is sufficient to assume that the transverse variability in the environment is mild enough that such scenarios may be neglected as exceptional cases rather than the norm.  The water column's bulk wavenumber field can theoretically vary in both the vertical $z$-direction and the transverse $y$-direction, but is predominantly characterized by a vertical convex profile while the transverse range-dependence is handled by invariant mappings induced by the adiabatic approximation.

The sea surface is an ideal horizontal planar pressure-release boundary at $z=0$.  The bathymetry profile varies in the transverse y-direction only and is interpolated as piecewise-linear for numerical convenience.  The acoustic properties of the sediment are also constrained to only transverse $y$-range-dependence, but are specified locally as a grazing-angle-dependent effective plane-wave reflection coefficient, $\FrR(y,\tht_b(y))$.  Even if the seafloor acoustic properties are the same at all locations, a particular ray-mode will in general have a unique set of $y$-dependent attenuation rates along its transverse cycle since the reflection coefficient depends on the incident wave's grazing angle which varies with the mapping induced by the adiabatic invariant.  Cumulative intensity attenuation due to bottom interaction is evaluated by performing a product integral over each unique transverse cycle trajectory once they are obtained via the cycle-tracking method.

Horizontal propagation in any direction within the waveguide is unbounded, but the two horizontal dimensions are handled differently so that a double mode-sum may be converted to an integration over solid angles.  The forward $x$-direction along which the wave cycle trajectories are parameterized imposes no restriction on propagation and simply satisfies the Sommerfeld radiation condition.  However, the transverse domain's solution is a mode summation which requires a finite domain on which to define the mode functions and their normalizations.  Artificial transparent boundaries are introduced with a specialized extinction distance reflection coefficient to simulate mode-stripping when modal energy escapes the computational domain of the waveguide.  This approach was first tested and verified in a minimal working example, the free-field Green's function energy flux model\supercite{langhirt2025foundations}.  Modes still contained within the model's transverse domain at a particular forward $x$-position contribute to the average energy flux within the entire waveguide, but each differential element of acoustic intensity is focused onto its corresponding mode’s horizontal cycle trajectory by the model's convergence factor kernel.  After specifying modenumbers for the vertical and transverse domains, the remaining forward component of the solution is solved as the impulse response of a source term that is fixed by the choice of modenumbers in the vertical and transverse domains.

\subsection{Notes on notation and conventions}

The 3D solid-angle energy flux model derived in this paper includes many interrelated and multivariate quantities that may be primarily dependent on one variable while other variables in the argument effectively function as parameters.  When an argument of a function has a vertical bar separating variables on the left and right, the function is understood to be primarily dependent on the variables to the left of the bar, and the variables to the right of the bar may be thought of as parameters; see \cref{eq:exampleVerticalBarArguments}).

\begin{equation}\label{eq:exampleVerticalBarArguments}
  G(\bvec{x}|\bvec{x}_0) \qq \text{or} \qq Z_m(z|y)
\end{equation}

The coordinates $(x,y,z)$ refer to a receiver location where the model is being evaluated at, and coordinates with a subscript zero $(x_0,y_0,z_0)$ refer to the source location.  When an evaluated quantity is subscripted with a zero and has no arguments, it implies that the quantity is being evaluated at the source location; see \cref{eq:exampleSourceQuantities}.

\begin{equation}\label{eq:exampleSourceQuantities}
  k_{z0} = k_z(z_0|y_0) \qq \text{or} \qq \IL_{y0} = \int_{y^{\lwr}}^{y_0}\frac{\dups}{k_y(\ups)}
\end{equation}

This energy flux model is based on integrating wavenumbers over complete and partial adiabatic ray cycles which may extend over either a portion of the computational domain or the entirety of the computational domain.  When a dummy coordinate is needed for a particular cycle integration, corresponding greek letters will be used; see \cref{eq:greekDummyCoords}.

\begin{equation}\label{eq:greekDummyCoords}
  \(x,y,z\) \q \leftrightarrow \q \(\xi,\ups,\zet\) 
\end{equation}

Minimum and maximum domain coordinates are denoted with superscript `$\min$' and `$\max$' respectively, whereas the lesser and greater turning points of a wave cycle are denoted with superscript `$\lwr$' and `$\upr$' respectively; see \cref{eq:minLwrDmyUprMax}.

\begin{equation}\label{eq:minLwrDmyUprMax}
  y^\min \le y^\lwr \le \ups \le y^\upr \le y^\max 
\end{equation}

When a quantity has a coordinate subscript, such as $k_z$ or $\IPhi_z$, then that quantity either exists within that coordinate domain of the problem, is oriented along that coordinate domain, or is computed as a cycle integration over that coordinate domain.  Note that $k_{xy}$ is the magnitude of the vector rejection of $\bvec{k}$ from $\bvec{k}_z$; see \cref{eq:kxyDefinition}.

\begin{equation}\label{eq:kxyDefinition}
  k_{xy}^2 = k^2 - k_z^2 = k_x^2 + k_y^2
\end{equation}

Quantities that are specific to a particular vertical or transverse mode are subscripted with the modenumbers $m$ and $n$ respectively; see \cref{eq:modenumberSubscripts}.  When a duplicate modenumber subscript is needed for projecting the partial differential equation (PDE) and its solution back onto the eigenbasis, or for computing the mode-sum cross-products, the secondary modenumber subscripts will have a hat accent.

\begin{equation}\begin{aligned}\label{eq:modenumberSubscripts}
  \text{vertical modes:}&\q Z_m \tand Z_{\hat{m}} \\
  \text{transverse modes:}&\q Y_{mn} \tand Y_{m\hat{n}}
\end{aligned}\end{equation}

The modenumbers are actually evaluated as cycle integrals since they are directly proportional to the phase integrals, $\wtil{\IPhi}_z$ and $\wtil{\IPhi}_y$, and for this purpose the double-struck glyphs $\Im$ and $\In$ are used instead; see \cref{eq:modenumberIntegrals}. After the energy flux solution is transformed from a mode-sum to an angular integration via the modal continuum approximation, the modenumbers can take on fractional values and no longer function as whole number indices, but they still identify the modes and uniquely map to propagation angles and wave trajectories.

\begin{equation}\begin{aligned}\label{eq:modenumberIntegrals}
  \text{vertical modenumber:}&\q \Im\pi = \wtil{\IPhi}_{zm}(y) = \int_{z^\lwr_m(y)}^{z^\upr_m(y)} k_{zm}(\zet|y)\dzet \\
  \text{transverse modenumber:}&\q \In\pi = \wtil{\IPhi}_{ymn} = \int_{y^{\lwr}_{mn}}^{y^{\upr}_{mn}} k_{ymn}(\ups)\dups
\end{aligned}\end{equation}

Quantities that are computed as cycle integrals are denoted with double-struck glyphs, and they include: the modenumbers $\Im$ and $\In$, the phase integrals $\IPhi_z$ and $\IPhi_y$, the cycle distances $\ID_z$ and $\ID_y$, the spectral cycle distances $\IL_z$ and $\IL_y$, and two convenience variables $\ILam$ and $\IChi$.  The cycle distances in particular should be interpreted carefully, e.g. the $\ID_z$ is the horizontal ($xy$-plane) distance that a vertical mode must travel for its phase to complete one vertical cycle from $z^\lwr$ to $z^\upr$ and back to $z^\lwr$; see \cref{eq:horzCycDistOfVertMode}.

\begin{equation}\label{eq:horzCycDistOfVertMode}
  \ID_z = 2k_{xy}\int_{z^\lwr}^{z^\upr}\frac{\dzet}{k_z(\zet)} 
\end{equation}

This energy flux model uses adiabatic ray cycles as scale-agnostic reference frames for tracking the evolution of the wave propagation within the waveguide.  Two cycle-integrated quantities are essential for this approach, the phase integral ($\IPhi$) and the spectral cycle distance ($\IL$), and they are calculated over both partial cycles and a complete half-cycle.  When the quantity is evaluated over a partial cycle, it is called an ``open'' cycle quantity, is denoted without accent, and includes an argument; see \cref{eq:exampleOpenCycleQuant}.  

\begin{equation}\label{eq:exampleOpenCycleQuant}
  \IPhi_y(y) = \int_{y^\lwr}^y k_y(\ups)\dups
\end{equation}

When the quantity is evaluated over a complete half-cycle, it is called a ``closed'' cycle quantity, is denoted with a tilde ($\sim$) accent, and excludes the positional argument; see \cref{eq:exampleClosedCycleQuant}.

\begin{equation}\label{eq:exampleClosedCycleQuant}
  \wtil{\IPhi}_y = \int_{y^\lwr}^{y^\upr} k_y(\ups)\dups 
\end{equation}

\subsection{Modal decomposition and the double mode-sum cross-product}\label{ssec:modalDecomp}

The model’s derivation begins with the 3D Helmholtz partial differential equation (PDE) for time-harmonic acoustic wave propagation, \cref{eq:initialHelmholtzEquation}, with a monopole source in three spatial dimensions as the inhomogeneous forcing term, corresponding to a solution as the 3D frequency-dependent Green's function subject to the environmental inhomogeneity and boundary conditions.

\begin{equation}
  \[\lap + k^2(z|y)\]p(\bvec{x}|\bvec{x_0}) = -S\delta(x)\delta(y-y_0)\delta(z-z_0) \label{eq:initialHelmholtzEquation} \\
\end{equation}

The first step in the analytical derivation is separating the vertical component of the problem, which requires adiabatically decoupling the range-dependence of the vertical modefunctions.  By assuming separability and substituting a vertical modesum for the pressure, \cref{eq:verticalModeExpansion}, and projecting the PDE back onto the vertical eigenbasis, \cref{eq:projectionOntoZEigenbasis}, coupling terms are revealed and subsequently neglected as insignificant.

\begin{gather}
  \Let p(\bvec{x}|\bvec{x_0}) = \sum_mA_m(x,y)Z_m(z) \tand \[\p[z][2] + k^2(z|y)\]Z_m(z|y) = +k_{xym}^2(y)Z_m(z|y) \label{eq:verticalModeExpansion} \\
  \int_0^{H(y)}Z_{\hat{m}}(z|y)\left\{\[\lap_{xy}+\p[z][2]+k^2(z|y)\]\sum_mA_m(x,y)Z_m(z|y) = -S\delta(x)\delta(y-y_0)\delta(z-z_0)\right\}\;\dz \label{eq:projectionOntoZEigenbasis} \\
  \Raro\q \[\lap_{xy}+k_{xym}^2(y)\]A_m(x,y) = -S\delta(x)\delta(y-y_0)Z_m(z_0|y) \label{eq:xyDomainPDE}
\end{gather}

The newly separated horizontal PDE for $A_m(x,y)$, \cref{eq:xyDomainPDE}, is also assumed to be separable and is expanded in a WKB eigenbasis for the transverse domain; see \cref{eq:transverseModeExpansion,eq:projectionOntoYEigenbasis}.  This time the adiabatic approximation is not required since the environment is assumed to be range-independent in the $x$-coordinate, and thus $k_{xmn}$ is spatially invariant.

\begin{gather}
  \Let A_m(x,y) = \sum_nX_{mn}(x)Y_{mn}(y) \tand \[\p[y][2]+k_{xym}^2(y)\]Y_{mn}(y) = {}^+k_{xmn}^2Y_{mn}(y) \label{eq:transverseModeExpansion} \\
  \int_{y_{mn}^\lwr}^{y_{mn}^\upr}Y_{m\hat{n}}(y)\left\{\[\p[x][2]+\p[y][2]+k_{xym}^2(y)\]\sum_nX_{mn}(x)Y_{mn}(y) = -S\delta(x)\delta(y-y_0)Z_m(z_0|y)\right\}\dy \label{eq:projectionOntoYEigenbasis} \\
  \Raro\qq \[\p[x][2]+k_{xmn}^2\]X_{mn}(x) = -S\delta(x)Y_{mn}(y_0)Z_m(z_0|y_0) \label{eq:xDomainODE}
\end{gather}

The final forward component of the problem, \cref{eq:xDomainODE}, has the form of a well-known inhomogeneous ODE with a standard fundamental solution, the free-field Green's function for the one-dimensional Helmholtz differential operator\supercite{arfken1999mathematical,haberman2012applied,morse1954methods}, \cref{eq:1DHelmholtzGreensFunction}.

\begin{equation}\label{eq:1DHelmholtzGreensFunction}
  X_{mn}(x) = \frac{iS}{2}Y_{mn}(y_0)Z_{m}(z_0|y_0)\frac{e^{ik_{xmn}\abs{x}}}{k_{xmn}} \\
\end{equation}

The complete expansion of the PDE's solution is now a double mode summation that represents the pressure field, and multiplying it by its complex conjugate yields the pressure amplitude squared, \cref{eq:pressureSquaredCrossProduct}.

\begin{multline}\label{eq:pressureSquaredCrossProduct}
  \CP = p^*p = \[\frac{iS}{2}\sum_{m=1}^{\IM}\sum_{n=1}^{\IN_m}\frac{e^{ik_{xmn}\abs{x}}}{k_{xmn}}Y_{mn}(y_0)Y_{mn}(y)Z_m(z_0|y_0)Z_m(z|y)\] \\ \times \[\frac{-iS}{2}\sum_{\hat{m}=1}^{\IM}\sum_{\hat{n}=1}^{\IN_m}\frac{e^{-ik_{x\hat{m}\hat{n}}\abs{x}}}{k_{x\hat{m}\hat{n}}}Y_{\hat{m}\hat{n}}(y_0)Y_{\hat{m}\hat{n}}(y)Z_{\hat{m}}(z_0|y_0)Z_{\hat{m}}(z|y)\]
\end{multline}

The double mode-sum cross-product in \cref{eq:pressureSquaredCrossProduct} can be represented as a sum of four terms by grouping together coherent and incoherent product pairings, producing a doubly-incoherent summation term, an $m$-coherent summation term, an $n$-coherent summation term, and a doubly-coherent summation term.  Conjugate symmetric exponential terms across the cross-product diagonals are combined into cosine terms using Euler's formula. Arguments of the modefunctions are temporarily dropped for notational compactness, and quantities denoted with a subscript `0' are evaluated at the source location instead of at the receiver.

\begin{align}
  \CP &= \CP_{II} + \CP_{CI} + \CP_{IC} + \CP_{CC} \label{eq:crossProductAsFourTerms} \\
  \CP_{II} &= \frac{S^2}{4}\sum_{m=1}^{\IM}\sum_{n=1}^{\IN_m} \frac{Y_{mn0}^2Y_{mn}^2Z_{m0}^2Z_m^2}{k_{xmn}^2} \label{eq:crossProductDoublyIncoherentTerm} \\
  \CP_{CI} &= \frac{S^2}{4}\sum_{m=2}^{\IM}\sum_{n=1}^{\IN_m}\sum_{\hat{m}=1}^{m-1}\frac{Y_{mn0}Y_{\hat{m}n0}Y_{mn}Y_{\hat{m}n}Z_{m0}Z_{\hat{m}0}Z_mZ_{\hat{m}}}{k_{xmn}k_{x\hat{m}n}}\times2\cos\(\(k_{xmn}-k_{x\hat{m}n}\)\abs{x}\) \label{eq:crossProductVerticalCoherentTerm} \\
  \CP_{IC} &= \frac{S^2}{4}\sum_{m=1}^{\IM}\sum_{n=2}^{\IN_m}\sum_{\hat{n}=1}^{n-1}\frac{Y_{mn0}Y_{m\hat{n}0}Y_{mn}Y_{m\hat{n}}Z_{m0}^2Z_{m}^2}{k_{xmn}k_{xm\hat{n}}}\times2\cos\(\(k_{xmn}-k_{xm\hat{n}}\)\abs{x}\) \label{eq:crossProductTransverseCoherentTerm} \\
  \CP_{CC} &= \frac{S^2}{4}\sum_{m=2}^{\IM}\sum_{n=2}^{\IN_m}\sum_{\hat{m}=1}^{m-1}\sum_{\hat{n}=1}^{n-1}\frac{Y_{mn0}Y_{\hat{m}\hat{n}0}Y_{mn}Y_{\hat{m}\hat{n}}Z_{m0}Z_{\hat{m}0}Z_{m}Z_{\hat{m}}}{k_{xmn}k_{x\hat{m}\hat{n}}}\times4\cos\(\(k_{xmn}-k_{x\hat{m}\hat{n}}\)\abs{x}\) \label{eq:crossProductDoublyCoherentTerm}
\end{align}

\subsection{WKB modefunctions and the cycle integrals}
\label{ssec:wkbModefunctionsAndCycleIntegrals}

Modefunctions of a convenient form are needed for substitution into the double mode-sum cross-product, \crefrange{eq:crossProductDoublyIncoherentTerm}{eq:crossProductDoublyCoherentTerm}, and adiabatic WKB modes are suitable for this purpose.  The WKB and modal continuum approximations allow full use of the ray-mode analogy and fluidity in interpreting individual propagating waves as either modes or rays.  Adiabatic modes use the concept of the reference waveguide\supercite{brekhovskikh2003fundamentals}, where horizontally-local eigenbases of vertical modefunctions $Z_m(z|y)$ are assumed to exist that solve \cref{eq:vertHelmholtzComponent} independently at each horizontal location.  

These vertical modefunctions, \cref{eq:vertWkbModefunctions}, are asymptotically approximated using the Wentzel-Kramers-Brillouin (WKB) method\supercite{bender1999advanced,brekhovskikh2003fundamentals}, but are assumed to be strictly real for energy flux methods and do not include the evanescent solutions outside of the vertical turning points $z_m^{\lwr}(y)$ and $z_m^{\upr}$.  The envelope of the modefunctions\supercite{smith1974averaged,zhou2013integrating} is given by \cref{eq:vertModeEnvelopes}, which represents the incoherent amplitude of the modefunctions with the oscillations averaged out.

\begin{gather}
  \[\p_z^2+k_{zm}^2(z|y)\]Z_m(z|y) = 0 \label{eq:vertHelmholtzComponent} \\
  k_{zm}(z|y) = \sqrt{k^2(z|y)-k_{xym}^2(y)} \label{eq:vertWavenumberComponent} \\
  Z_m(z|y) \sim \sqrt{\frac{2}{k_{zm}(z|y)\wtil{\IL}_{zm}(y)}}\cos(\IPhi_{zm}(z|y)) \label{eq:vertWkbModefunctions} \\
  \overline{\abs{Z_m(z|y)}^2} = \frac{1}{k_{zm}(z|y)\wtil{\IL}_{zm}(y)}  \label{eq:vertModeEnvelopes}
\end{gather}

Each vertical mode has a unique associated modenumber, $\Im$, directly proportional to the closed phase integral, \cref{eq:vertClosedPhaseIntegral}.  The maximum possible vertical modenumber at any horizontal location is associated with a vertically propagating wave bouncing between the sea surface and seafloor at that location, \cref{eq:vertMaxModenumber}.  Each wave's modenumber is held spatially invariant by the adiabatic approximation which induces an effective horizontal wavenumber profile, $k_{xym}(y)$, that is unique to each vertical mode, \cref{eq:vertWavenumberComponent}.

\begin{gather}
  \IPhi_{zm}(z|y) = \int_{z_m^\lwr(y)}^{z}k_{zm}(\zet|y)\dzet \label{eq:vertOpenPhaseIntegral} \\
  \wtil{\IPhi}_{zm}(y) = \int_{z_m^\lwr(y)}^{z_m^\upr(y)}k_{zm}(\zet|y)\dzet = \Im\pi \label{eq:vertClosedPhaseIntegral} \\
  \IM(y) = \frac{1}{\pi}\int_0^{H(y)}k(\zet|y)\dzet \label{eq:vertMaxModenumber}
\end{gather}

The vertical modefunctions have a horizontal cycle distance, $\ID_z$, that is also calculated by integrating over the vertical cycle.  For the purposes of this derivation, it is simpler to work with a quantity that may be called the ``spectral cycle distance,'' borrowing a convention from optics where the modifier ``spectral'' means that the quantity has been multiplied or divided by a frequency or wavelength factor.

\begin{gather}
  \IL_{zm}(z|y) = \int_{z_m^\lwr(y)}^{z}\frac{\dzet}{k_{zm}(\zeta|y)} \label{eq:vertOpenSpectralCycDist} \\
  \wtil{\IL}_{zm}(y) = \int_{z_m^\lwr(y)}^{z_m^\upr(y)}\frac{\dzet}{k_{zm}(\zet|y)} = \frac{\ID_{zm}(y)}{2k_{xym}(y)} \label{eq:vertClosedSpectralCycDist}
\end{gather}

Both the vertical phase integrals (\cref{eq:vertOpenPhaseIntegral,eq:vertClosedPhaseIntegral}) and the vertical spectral cycle distances (\cref{eq:vertOpenSpectralCycDist,eq:vertClosedSpectralCycDist}) have open and closed forms.  The open integrals take a positional argument as one of the integration limits, whereas the closed integrals are integrated over the complete half-cycle from one turning point to the other\supercite{harrison2013ray,langhirt2025foundations}.

Analogous versions of these quantities and expressions also exist for the transverse modal expansion.  The transverse eigenvalue problem is given by \cref{eq:transHelmholtzComponent}, and the transverse wavenumber component is given by \cref{eq:transWavenumberComponent}.  The transverse WKB approximated modefunctions are given by \cref{eq:transWkbModefunctions}, and the mode envelopes are given by \cref{eq:transModeEnvelopes}.

\begin{gather}
    \[\p_y^2+k_{ymn}^2(y)\]Y_{mn}(y) = 0 \label{eq:transHelmholtzComponent} \\
    k_{ymn}(y) = \sqrt{k_{xym}^2(y)-k_{xmn}^2} \label{eq:transWavenumberComponent} \\
    Y_{mn}(y) \sim \sqrt{\frac{2}{k_{ymn}(y)\wtil{\IL}_{ymn}}}\cos(\IPhi_{ymn}(y)) \label{eq:transWkbModefunctions} \\
    \overline{\abs{Y_{mn}(y)}^2} = \frac{1}{k_{ymn}(y)\wtil{\IL}_{ymn}} \label{eq:transModeEnvelopes}
\end{gather}

The transverse open phase integral is given by \cref{eq:transOpenPhaseIntegral}, and the close form is given by \cref{eq:transClosedPhaseIntegral}.  The maximum transverse modenumber is given by \cref{eq:transMaxModenumber}.

\begin{gather}
    \IPhi_{ymn}(y) = \int_{y_{mn}^\lwr}^{y}k_{ymn}(\ups)\dups \label{eq:transOpenPhaseIntegral} \\
    \wtil{\IPhi}_{ymn} = \int_{y_{mn}^\lwr}^{y_{mn}^\upr}k_{ymn}(\ups)\dups = \In\pi \label{eq:transClosedPhaseIntegral} \\
    \IN_m = \frac{1}{\pi}\int_{y^\min}^{y^\max}\Real{k_{xym}(\ups)}\dups \label{eq:transMaxModenumber}
\end{gather}

The transverse open spectral cycle distance is given by \cref{eq:transOpenSpectralCycDist}, and the closed form is given by \cref{eq:transClosedSpectralCycDist}

\begin{gather}
    \IL_{ymn}(y) = \int_{y_{mn}^\lwr}^y\frac{\dups}{k_{ymn}(\ups)} \label{eq:transOpenSpectralCycDist} \\
    \wtil{\IL}_{ymn} = \int_{y_{mn}^\lwr}^{y_{mn}^\upr}\frac{\dups}{k_{ymn}(\ups)} = \frac{\ID_{ymn}}{2k_{xmn}} \label{eq:transClosedSpectralCycDist}
\end{gather}

The normalized WKB modefunctions, \cref{eq:vertWkbModefunctions,eq:transWkbModefunctions}, are expressed as exponential or trigonometric functions of the phase integrals that are amplitude modulated by the cycle distances.  The phase integrals and cycle distances are closely related in the WKB approximation, and the derivative of the modal eigenvalue with respect to the modenumber (the mode separation) can be expressed in terms of the cycle distance\supercite{milder1969ray,brekhovskikh2003fundamentals,harrison2013ray}.

Returning to the doubly incoherent pressure-squared term, \cref{eq:crossProductDoublyIncoherentTerm}, and substituting in the WKB modal-amplitude-squared envelopes yields a discrete mode-sum that is proportional to the locally averaged acoustic intensity and expressed solely in terms of wavenumber components and cycle integrals.  Assuming modal continuums in both $z$ and $y$ implies the existence of dense sets of propagating finite-frequency modes, which is effectively a high-frequency approximation in relation to the depth and width of the waveguide.  Regardless of whether one considers a limiting case towards infinite frequency in a finite depth waveguide or finite frequency in an infinite depth waveguide, there is assumed to be a continuum of trapped and leaky modes spanning the propagation angles from $0\deg$ to $90\deg$.  Then the summations may be converted to integrals and the modenumber differences become differential mode number elements, as shown in \cref{eq:doublyIncoherentIntegralOverModenumbers}.

\begin{equation}\begin{split}\label{eq:doublyIncoherentIntegralOverModenumbers}
    \CP &= \frac{S^2}{4}\int_0^{\IM}\int_0^{\IN_m}\frac{\d\Im\d\In}{k_{x^2}k_{y0}k_{y}k_{z0}k_{z}\wtil{\IL}_{y}^2\wtil{\IL}_{z}\wtil{\IL}_{z0}} \\
\end{split}\end{equation}

\subsection{The model's differential topology, differential chains, and Jacobian}\label{ssec:modelsDifferentialTopologyDifferentialChainsAndJacobian}

In order to derive an appropriate Jacobian transformation and suitable Taylor approximations of the inter-modal interference, it is necessary to first carefully consider the model's differential topology.  The general spatially-dependent propagation angles that are mapped by invariance relations, $\(\tht(z|y),\phi(y)\)$, are conceptually distinct from the initial propagation angles defined at the source, $\(\tht_0,\phi_0\)$.  

Under the adiabatic approximation, the modenumbers $\Im$ and $\In$ are invariant throughout the waveguide and uniquely identify individual modes.  Each $(m,n)$-mode has modal eigenvalues $k_{xym}(y)$ and $k_{xmn}$, and each eigenvalue pair corresponds to a mapping of propagation angles $\theta_{m}(z|y)$ (elevation) and $\phi_{mn}(y)$ (azimuth) along the wave's unique trajectory identified by the source take-off angles $\tht_0=\tht(z_0,y_0)$ and $\phi_0=\phi(y_0)$.

This implies the existence of a common reference manifold, $\CM\in\IR^4$ with a canonical basis of the model's independent parameters $\{y,z,\tht_0,\phi_0\}$, from which all other variables in the model may be considered homotopically equivalent to an immersed submanifold of $\CM$. Immersions that are further constrained to be structure-preserving injective mappings are called embeddings, and two differentiable variables that are homeomorphic to the same embedded submanifold of $\CM$ are implicitly diffeomorphic to each other.

Derivatives need only be defined as differentials of a quantity with respect to differentials of the independent parameters, so the only explicit derivatives that need to be defined are those that traverse these relations in reverse order back to the independent parameters.  Any two subsets of variables that are diffeomorphic to the same embedded submanifold of $\CM$ must also be diffeomorphic to each other.  Only derivatives between diffeomorphic submanifolds are considered invertible, and since all manifolds of variables in the model are diffeomorphic to submanifolds of $\CM$, then derivatives between any two diffeomorphic manifolds in the model can be defined through a submanifold of $\CM$ by inverting either one of their total derivatives, i.e. the Jacobian matrices.

The modenumbers, $\(\Im,\In\)$, or closed phase integrals, $\(\IPhi_z,\IPhi_y\)$, are diffeomorphic to the initial propagation angles, $\(\tht_0,\phi_0\)$.  At any fixed receiver position, $(y,z)$, the open phase integrals, $(\IPhi_z(z|y),\IPhi_y(y))$, are diffeomorphic to submanifolds of $(\tht_0,\phi_0)$.  Then a multivariate change of variables between these manifolds is uniquely defined by invertible linear transformations of differentials.

\begin{figure}[!htb]
  \includegraphics[width=\textwidth]{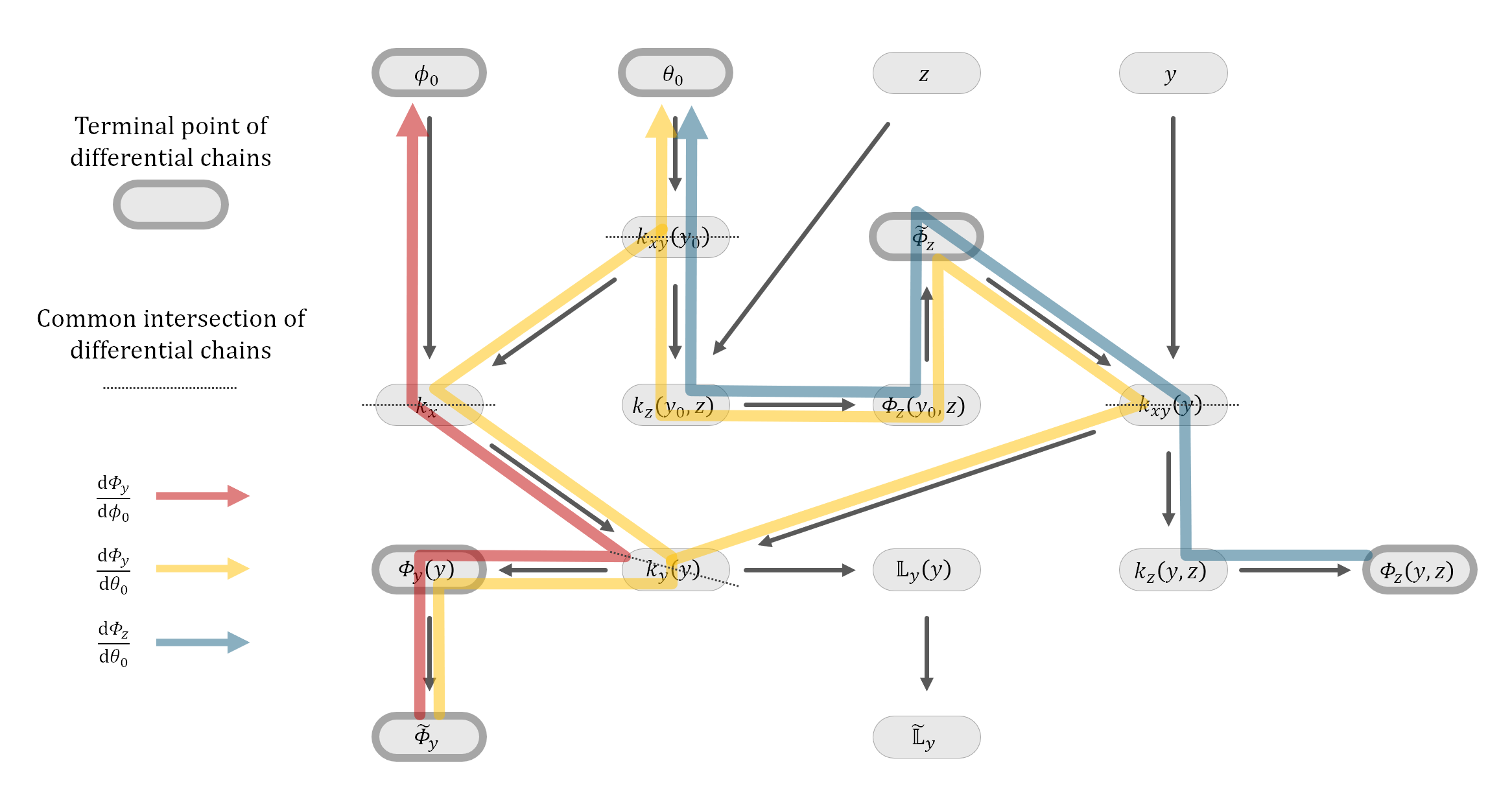}
  \caption{Univariate differential chains between the cycle integrals and the model's independent parameters}
  \label{fig:DifferentialChains}
\end{figure}

\Cref{fig:DifferentialChains} shows the flow of calculations (black arrows) beginning from the model's canonical basis of parameters (top row).  It should be noted that the relation deriving $k_{xy}(y)$ from $\wtil{\IPhi}_z$ is implicitly defined by the vertical modenumber invariance of the adiabatic modes approximation.  The closed and open phase integrals are the quantities of interest that need their differentials mapped back to the source propagation angles.  The univariate differential chains that flow from a single phase integral to a single source propagation angle are depicted as colored arrows tracing back up in the opposite direction: $\d\IPhi_y/\d\phi_0$ is red, $\d\IPhi_y/\d\theta_0$ is yellow, and $\d\IPhi_z/\d\theta_0$ is blue.  There is no differential chain for $\d\IPhi_z/\d\phi_0$ because the vertical mode expansion is defined independently of azimuthal angles and before the transverse modes are introduced, hence $\d\IPhi_z/\d\phi_0=0$ always.  Using the differential chains illustrated in \cref{fig:DifferentialChains}, multivariate coordinate transformations can be defined that are topologically consistent under the model's assumptions and constraints.  The necessary intermediary univariate derivatives are defined and evaluated in \crefrange{eq:dkxy0dth0}{eq:dPhiydkx}.

\begin{align}
    k_{xy0} &= k_0\cos\tht_0 &\Raro&& \dd{k_{xy0}}{\tht_0} &= -k_0\sin\tht_0 = -k_{z0} \label{eq:dkxy0dth0} \\ \spacedline{5}
    k_z(z|y) &= \sqrt{k^2(z|y)-k_{xy}^2(y)} &\Raro&& \dd{k_z}{k_{xy}} &= \frac{-k_{xy}}{k_z} \label{eq:dkzdkxy} \\
    &&\Raro&& \dd{k_z(z|y_0)}{k_{xy0}} &= \frac{-k_{xy0}}{k_z(z|y_0)} \label{eq:dkzdkxy0} \\ \spacedline{5}
    \wtil{\IPhi}_z(y) &= \int_{z^\lwr(y)}^{z^\upr(y)}k_z(\zet|y)\dzet &\Raro&& \dd{\wtil{\IPhi}_z(y)}{k_{xy}} &= -k_{xy}\wtil{\IL}_{z} \label{eq:dTPhizdkxy} \\
    &&\Raro&& \dd{\wtil{\IPhi}_{z0}}{k_{xy0}} &= -k_{xy0}\wtil{\IL}_{z0} \label{eq:dTPhiz0dkxy0} \\ \spacedline{5}
    \IPhi_z(z|y) &= \int_{z^\lwr(y)}^zk_z(\zet|y)\dzet &\Raro&& \dd{\IPhi_z(z|y)}{k_{xy}} &= -k_{xy}\IL_{z} \label{eq:dPhizdkxy} \\ \spacedline{5}
    \evalat{\wtil{\IPhi}_z}{y_0} &= \evalat{\wtil{\IPhi}_z}{y} &\overset{\ndd{}{\tht_0}}{\Raro}&& \dd{k_{xy}}{k_{xy0}} &= \frac{k_{xy0}\wtil{\IL}_{z0}}{k_{xy}\wtil{\IL}_{z}} \label{eq:dkxydkxy0} \\ \spacedline{5}
    k_x &= k_{xy0}\cos\phi_0 &\Raro&& \pp{k_x}{k_{xy0}} &= \cos\phi_0 = \frac{k_x}{k_{xy0}} \label{eq:dkxdkxy0} \\
    &&\Raro&& \pp{k_x}{\phi_0} &= -k_{xy0}\sin\phi_0 = -k_{y0} \label{eq:dkxdphi0} \\ \spacedline{5}
    k_y(\ups) &= \sqrt{k_{xy}^2(\ups)-k_x^2} &\Raro&& \pp{k_y}{k_{xy}} &= \frac{k_{xy}}{k_y} \label{eq:dkydkxy} \\
    &&\Raro&& \pp{k_y}{k_x} &= \frac{-k_x}{k_y} \label{eq:dkydkx} \\ \spacedline{5}
    \wtil{\IPhi}_y &= \int_{y^\lwr}^{y^\upr}k_y(\ups)\dups &\Raro&& \pp{\wtil{\IPhi}_y}{k_{xy}} &= \int_{y^\lwr}^{y^\upr}\frac{k_{xy}(\ups)}{k_y(\ups)}\dups \label{eq:dTPhiydkxy} \\
    &&\Raro&& \pp{\wtil{\IPhi}_y}{k_x} &= -k_x\wtil{\IL}_y \label{eq:dTPhiydkx} \\ \spacedline{5}
    \IPhi_y(y) &= \int_{y^\lwr}^yk_y(\ups)\dups &\Raro&& \pp{\IPhi_y}{k_{xy}} &= \int_{y^\lwr}^y\frac{k_{xy}(\ups)}{k_y(\ups)}\dups \label{eq:dPhiydkxy} \\
    &&\Raro&& \pp{\IPhi_y}{k_x} &= -k_x\IL_y \label{eq:dPhiydkx}
\end{align}

Using these derivatives and following the differential chains depicted in \cref{fig:DifferentialChains}, the total directional derivatives of the phase integrals along differentials of the independent parameters $\tht_0$ and $\phi_0$ may be constructed.  New convenience notation is introduced in \cref{eq:phaseIntegralConvenienceNotation} to keep the expressions compact.  In some cases, taking derivatives of functionals (the cycle integrals) requires using Leibniz differentiation under the integral. Alternatively they can be evaluated more rigorously using functional calculus, but careful consideration of the differential topology shows that the differentials may be simplified to standard techniques of multivariate calculus.

\begin{gather}\label{eq:phaseIntegralConvenienceNotation}
    \Let \left\{\q \begin{gathered}\tau_{z} = \tan\tht = \frac{k_{z}}{k_{xy}} \tand \tau_{y} = \tan\phi = \frac{k_{y}}{k_x} \\
    \wtil{\ILam}_y = k_x^2\wtil{\IL}_y \comma \wtil{\ILam}_z(y) = k_{xy}^2(y)\wtil{\IL}_{z}(y) \command \wtil{\ILam}_{z0} = k_{xy0}^2\wtil{\IL}_{z0} \\
    \ILam_y(y) = k_x^2\IL_y(y) \tand \ILam_z(z|y) = k_{xy}^2(y)\IL_z(z|y) \\
    \wtil{\IL}_{yz} = \int_{y^\lwr}^{y^\upr}\frac{1}{k_y(\ups)\wtil{\IL}_z(\ups)}\dups \tand \IL_{yz}(y) = \int_{y^\lwr}^{y}\frac{\dups}{k_y(\ups)\wtil{\IL}_z(\ups)} \end{gathered} \right.
\end{gather}

\begin{equation}\begin{aligned}
  \d\wtil{\IPhi}_z &= \int_{z^\lwr(y)}^{z^\upr(y)} \pp{k_z}{k_{xy}}\pp{k_{xy}}{k_{xy0}}\pp{k_{xy0}}{\tht_0}\dzet\d\tht_0 + \cancelto{0}{\[k_z(\zet|y)\pp{\zet}{\tht_0}\]_{z^\lwr(y)}^{z^\upr(y)}}\d\tht_0 + \cancelto{0}{\p_{\phi_0}\wtil{\IPhi}_z}\dphi_0 \\
  &= \tau_{z0}\wtil{\ILam}_{z0}\dtht_0 + 0\dphi_0
\end{aligned}\end{equation}

\begin{equation}\begin{aligned}
  \d\wtil{\IPhi}_y &= \cancelto{0}{\[k_y(\ups)\pp{\ups}{\tht_0}\]_{y^\lwr}^{y^\upr}}\dtht_0 + \int_{y^\lwr}^{y^\upr}\[\pp{k_y}{k_{xy}}\pp{k_{xy}}{k_{xy0}}+\pp{k_y}{k_x}\pp{k_x}{k_{xy0}}\]\pp{k_{xy0}}{\tht_0}\dups\dtht_0 \\
  &\q + \cancelto{0}{\[k_y(\ups)\pp{\ups}{\phi_0}\]_{y^\lwr}^{y^\upr}}\dphi_0 + \int_{y^\lwr}^{y^\upr}\pp{k_y}{k_x}\pp{k_x}{\phi_0}\dups\dphi_0 \\
  &= \tau_{z0}\(\wtil{\ILam}_y-\wtil{\ILam}_{z0}\wtil{\IL}_{yz}\)\dtht_0 + \tau_{y0}\wtil{\ILam}_y\dphi_0
\end{aligned}\end{equation}

%

The closed phase integrals are directly proportional to the modenumbers by a constant factor of $\pi$, so their Jacobian, \cref{eq:closedPhaseIntegralJacobian}, and its inverse, \cref{eq:inverseClosedPhaseIntegralJacobian}, are used as a Jacobian determinant, \cref{eq:modenumberJacobianDeterminant}, to transform between differentials on the $(\Im,\In)$-manifold and differentials on the $(\tht_0,\phi_0)$-manifold.

\begin{align}
    \bmat{\d\Im \\ \d\In} &= \pi^{-1}\pmat{\tau_{z0}\wtil{\ILam}_{z0} & 0 \\ \tau_{z0}\(\wtil{\ILam}_y-\wtil{\ILam}_{z0}\wtil{\IL}_{yz}\) & \tau_{y0}\wtil{\ILam}_y}\bmat{\d\tht_0 \\ \d\phi_0} = \(\frac{\D(\Im,\In)}{\D(\tht_0,\phi_0)}\)\bmat{\d\tht_0 \\ \d\phi_0} \label{eq:closedPhaseIntegralJacobian} \\
    \bmat{\d\tht_0 \\ \d\phi_0} &= \pi\pmat{\[\tau_{z0}\wtil{\ILam}_{z0}\]^{-1} & 0 \\[6pt] \frac{\wtil{\ILam}_{z0}\wtil{\IL}_{yz}-\wtil{\ILam}_y}{\tau_{y0}\wtil{\ILam}_y\wtil{\ILam}_{z0}} & \[\tau_{y0}\wtil{\ILam}_y\]^{-1}}\bmat{\d\Im \\ \d\In} = \(\frac{\D(\tht_0,\phi_0)}{\D(\Im,\In)}\)\bmat{\d\Im \\ \d\In} \label{eq:inverseClosedPhaseIntegralJacobian} \\
    \d\Im\d\In &= \abs{\frac{\D(\Im,\In)}{\D(\tht_0,\phi_0)}}\d\tht_0\d\phi_0 = \pi^{-2}\tau_{z0}\tau_{y0}\wtil{\ILam}_y\wtil{\ILam}_{z0}\d\tht_0\d\phi_0 = \pi^{-2}k_xk_{y0}k_{z0}k_{xy0}\wtil{\IL}_y\wtil{\IL}_{z0}\d\tht_0\d\phi_0 \label{eq:modenumberJacobianDeterminant}
\end{align}

The Jacobian for the open phase integrals is defined and evaluated similarly, yielding \cref{eq:openPhaseIntegralJacobian}.

\begin{equation}\label{eq:openPhaseIntegralJacobian}
  \bmat{\d\IPhi_z(z|y) \\ \d\IPhi_y(y)} = \pmat{\tau_{z0}\frac{\wtil{\ILam}_{z0}}{\wtil{\ILam}_z}\ILam_z & 0 \\ \tau_{z0}\(\ILam_y-\wtil{\ILam}_{z0}\IL_{yz}\) & \tau_{y0}\ILam_y} \bmat{\d\tht_0 \\ \d\phi_0}
\end{equation}


Derivation of the convergence factor will require an expression for the total derivative of the open phase integrals with respect to the modenumbers, \cref{eq:openClosedPhaseIntegralJacobian}, which is the product of the Jacobians from \cref{eq:openPhaseIntegralJacobian} and \cref{eq:inverseClosedPhaseIntegralJacobian}.  Two new symbols are introduced for notational convenience in \cref{eq:fractionalCyclePhaseOffsetDefinition}, which are the cycle phase offsets for the vertical and transverse cycles expressed as fractional cycles or ``turns'' instead of radians.  The cross-domain spectral cycle distances, $\IL_{yz}$ and $\wtil{\IL}_{yz}$, are accumulating the elapsed vertical cycles per partial or total transverse cycle respectively, and thus effectively function as a cross-domain correction to the cycle phase offsets due to the interdependence of the double mode summation.  Changes in the modal interference at some fixed $(y,z)$-position due to varying the ray's initial propagation angles, $(\tht_0,\phi_0)$, can be approximated using the vertical cycle phase, transverse cycle phase, and this cross-domain proportion of elapsed cycles accumulated along the wave's horizontal trajectory.  The physical meaning of the differential transformation in \cref{eq:openClosedPhaseIntegralJacobian} is that small changes in modenumbers, $\d\Im$ and $\d\In$, can be converted to small changes in modal phases, $\d\IPhi_z$ and $\d\IPhi_y$, for all $(y,z)$-positions within the environment.

\begin{equation} \label{eq:fractionalCyclePhaseOffsetDefinition}
  \Let \IChi_y(y) = \frac{\IL_y(y)}{\abs{\wtil{\IL}_y}} = \frac{\ILam_y(y)}{\abs{\wtil{\ILam}_y}} \tand \IChi_z(z|y) = \frac{\IL_z(z|y)}{\abs{\wtil{\IL}_z(y)}} = \frac{\ILam_z(z|y)}{\abs{\wtil{\ILam}_z(y)}} \\
\end{equation}

\begin{gather}\begin{aligned}\label{eq:openClosedPhaseIntegralJacobian}
    \bmat{\d\IPhi_z(z|y)\\\d\IPhi_y(y)} &= \DD{(\IPhi_z,\IPhi_y)}{(\tht_0,\phi_0)}\[\DD{(\Im,\In)}{(\tht_0,\phi_0)}\]^{-1}\bmat{\dtht_0\\\dphi_0} \\
    &= \pi\pmat{ \IChi_z & 0 \\[6pt] \IChi_y\wtil{\IL}_{yz}-\IL_{yz} & \IChi_y }\bmat{\d\Im \\ \d\In}
\end{aligned}\end{gather}

The convergence factor derivation also uses one other derivative of the forward wavenumber component with respect to the modenumbers, provided here in \cref{eq:dkxdmn}.

\begin{gather}\begin{aligned}\label{eq:dkxdmn}
    \bmat{\dk_x} &= \frac{-\pi k_x}{\wtil{\ILam}_y}\pmat{\wtil{\IL}_{yz} & 1}\bmat{\d\Im \\ \d\In}
\end{aligned}\end{gather}

\subsection{Unfolding the lossless incoherent solid-angle energy flux integral and limiting the WKB amplitudes}\label{ssec:unfoldingLosslessIncoherentSolidAngleEnergyFluxIntegralLimitingWkbAmplitudes}

The Jacobian determinant, \cref{eq:modenumberJacobianDeterminant}, can be substituted into the doubly-incoherent pressure-squared term from the mode-sum cross-product, \cref{eq:doublyIncoherentIntegralOverModenumbers}, to obtain a fully incoherent solution of the acoustic field in a tube of constant cross section, \cref{eq:doublyIncoherentFolded3DSAEF}.

\begin{equation}\begin{aligned}\label{eq:doublyIncoherentFolded3DSAEF}
    \CP_{II} &= \frac{S^2}{4\pi^2}\int_0^{\pi/2}\int_0^{\pi/2}\frac{k_{xy0}}{k_xk_yk_z\wtil{\IL}_y\wtil{\IL}_z}\dtht_0\dphi_0
\end{aligned}\end{equation}

This solution is not particularly useful or realistic for ocean acoustic problems since it does not contain any loss mechanisms due to bottom interactions, no mechanism for stripping transverse modal energy from the computational domain to simulate geometric spreading, or the approximations of neighboring mode interference that produce convergence and shadow zones.  Since acoustic energy is conserved within this lossless duct, this solution is analogous to an incoherent 1D Helmholtz free-field Green's function.  By conservation of energy, the averge acoustic energy density propagating along a tube will remain the same if it cannot geometrically spread and there are no dissipative or attenuative processes along the boundaries or within the volume.  However it is an important intermediary solution upon which the 3D solid-angle energy flux model is built, and it demonstrates an acoustic field solution over $(x,y,z)$ computed by a solid-angle integration of distributions of differential acoustic intensity emanating from the source.

The partial and total cycle integrals have been defined to preserve wave propagation orientation, but the model has not yet utilized this asymmetry.  Even though the negative and positive angular domains are currently symmetric, the integrations over propagation angle can be unfolded to illustrate how the energy flux integral will incorporate asymmetric kernels.  The loss-less pressure-squared is directly proportional to the acoustic energy flux density which is a conserved quantity.  So when an angular domain is unfolded, $\(0,\tfrac{\pi}{2}\)\negmedspace\to\negmedspace\(\tfrac{-\pi}{2},\tfrac{+\pi}{2}\)$, then the differential pressure-squared contribution represented by the integrand should be contravariantly halved to conserve the total acoustic power emanating from the source.  Then if the TL-normalized monopole amplitude is substituted, $S=4\pi$, all of the constant factors neatly cancel each other out as shown in \cref{eq:doublyIncoherentUnfolded3DSAEF}.

\begin{equation}\label{eq:doublyIncoherentUnfolded3DSAEF}
    \CP_{II} = \int_{\nfrac{-\pi}{2}}^{\nfrac{+\pi}{2}}\int_{\nfrac{-\pi}{2}}^{\nfrac{+\pi}{2}}\frac{k_{xy0}}{k_xk_yk_z\wtil{\IL}_y\wtil{\IL}_z}\dtht_0\dphi_0 \\
\end{equation}

The WKB modal amplitudes expressed in \cref{eq:doublyIncoherentUnfolded3DSAEF} are singular at their turning points, as may be seen by letting $k_y\raro0$ or $k_z\raro0$.  The full formal derivation of WKB modefunctions often includes Airy function approximations in the vicinity of the turning points, which are $k^2$-linear approximations asymptotically-matched between the modefunctions' outer evanescent solution and inner oscillating solution.  Harrison proposed using a simple WKB-limiter mechanism to cap the modal amplitudes to a maximum that roughly corresponds to the Airy function solutions\supercite{harrison2013ray}.  This modification is provided here in \cref{eq:wkbLimiterModification}.

\begin{align}
    \Let &\left\{\;\begin{aligned} \overline{\abs{Y}^2} &= \frac{1}{\hat{k}_y\wtil{\IL}_y} \twhere \hat{k}_y = \opmax\(k_y\;,\; \sqrt[3]{3\pi k_{xy}\abs{\dd{k_{xy}}{y}}}\) \\
      \overline{\abs{Z}^2} &= \frac{1}{\hat{k}_z\wtil{\IL}_z} \twhere \hat{k}_z = \opmax\(k_z\;,\; \sqrt[3]{3\pi k\abs{\dd{k}{z}}}\) \end{aligned} \right. \label{eq:wkbLimiterModification} \\
    \CP_{II} &= \int_{\nfrac{-\pi}{2}}^{\nfrac{+\pi}{2}}\int_{\nfrac{-\pi}{2}}^{\nfrac{+\pi}{2}}\frac{k_{xy0}}{k_x\hat{k}_y\hat{k}_z\wtil{\IL}_y\wtil{\IL}_z}\dtht_0\dphi_0 \label{eq:wkbLimitedIncoherent3DSAEF}
\end{align}

The integrand from \cref{eq:wkbLimitedIncoherent3DSAEF} is the modal amplitude-squared kernel, \cref{eq:modalAmplitudeSquaredKernel}, and represents the base incoherent solution that will be further modulated and attenuated by additional kernels.

\begin{equation} \label{eq:modalAmplitudeSquaredKernel}
  \Psi = \frac{k_{xy0}}{k_x\hat{k}_y\hat{k}_z\wtil{\IL}_y\wtil{\IL}_z}
\end{equation}

This integral can be evaluated on a custom angular domain for various reasons, \cref{eq:incoherentEnergyFluxIntegralWithCustomAngularDomain}.  There is no analytical reason for restricting the domain of azimuthal angles since they are associated with an unbounded continuous spectrum in an almost horizontally-stratified waveguide, but $0\deg$ and $90\deg$ can sometimes cause numerical instabilities in the calculations, so a small angular epsilon buffer, $\veps_\phi$, can be defined to avoid integrating over these angles, \cref{eq:customSourcePhiDomain}.  In a lossy waveguide with penetrable bottom, the steeper elevation angles beyond the critical angle will decay rapidly and contribute less to the far-field intensity, so the elevation upper bound may be set to a user-specified limit, see $\abs{\tht}_\max$ in \cref{eq:customSourceThetaDomain}.  The energy flux solution is inherently a form of high-frequency approximation leveraging the ray-mode analogy, and as such there is no inherent physical reason for a minimum elevation angle cutoff.  In order to clearly see the effects of horizontal refraction, a minimum elevation angle magnitude, $\abs{\tht}_\min$, can be set that corresponds to the finite-frequency mode-cutoff wavenumber\supercite{harrison1979acoustic}.

\begin{align}
  \Omg_\tht &= \{\tht_0\in\[\ngtv\abs{\tht}_\max,\ngtv\abs{\tht}_\min\]\cup\[\pstv\abs{\tht}_\min,\pstv\abs{\tht}_\max\]\} \label{eq:customSourceThetaDomain} \\
  \Omg_\phi &= \left\{\phi_0\in\[\ngtv\tfrac{\pi}{2}+\veps_\phi,\ngtv\veps_\phi\]\cup\[\pstv\veps_\phi,\pstv\tfrac{\pi}{2}-\veps_\phi\]\right\} \label{eq:customSourcePhiDomain} \\
  \CP &= \int_{\Omg_\tht}\int_{\Omg_\phi}\Psi\dphi_0\dtht_0 \label{eq:incoherentEnergyFluxIntegralWithCustomAngularDomain}
\end{align}

An example finite-frequency cutoff angle is provided in \cref{eq:idealWaveguideCutoffAngle} that is based on the maximum allowable vertical wavelength in an ideal waveguide, \cref{eq:idealModalCutoffWavelength}.

\begin{align}
  \lam_{z}^\max &= 4H_0 \label{eq:idealModalCutoffWavelength} \\
  \abs{\tht}_\min &= \asin\(\frac{c_0}{4fH_0}\) \label{eq:idealWaveguideCutoffAngle}
\end{align}

There are three additional mechanisms to add to this model that will produce the appropriate longitudinal $x$-dependence of the acoustic field solution: the bivariate convergence factor, the transverse mode-stripping kernel, and the bottom attenuation kernel.

\subsection{Derivation of the bivariate convergence factor from the double mode-sum cross-product}\label{ssec:derivationBivariateConvergenceFactorDoubleModeSumCrossProduct}

The convergence factor is derived as an approximation of neighboring modal interference, and it effectively acts like an eigenray filter that captures convergence and shadow zone features.  Incorporating the additional coherent terms without introducing loss mechanisms should modulate the incoherent acoustic field without changing the net energy flux density transmitted through the waveguide's ($y,z$)-cross-section.  The convergence factor derivation for this 3D model closely follows Harrison's derivation of the pre-summed cosine convergence factor in his 2D semi-coherent energy flux model\supercite{harrison2013ray}, however the bivariate convergence factor derived for this model is more complex due to nested mode expansions in the cross-product, cross-domain inter-modal interference, and the need to distinguish between opposite wave orientations to break the energy flux integral's angular symmetry.

One might expect that vertical coherence resolves vertical refraction and convergence while the transverse coherence resolves transverse refraction and convergence. This is the conceptual motivation for the model and a useful approximate description of how it works, but it is a coarse simplification of the model's behavior.  Correspondences between these cross-product terms is better understood by conceptually distinguishing between (a) differences in modenumbers, $\D(\Im,\In)$, and (b) differences in modal phase $\D(\IPhi_z,\IPhi_y)$.  Altogether there is a transverse $\In$-coherence that produces only transverse $\IPhi_y$-interference, a vertical $\Im$-coherence that produces vertical $\IPhi_z$-interference and modifies the transverse $\IPhi_y$-interference, and a coupled interaction between vertical $\Im$-coherence and transverse $\In$-coherence that modifies both the vertical ($\IPhi_z$) and transverse ($\IPhi_y$) interferences.

In order for the cycle-tracking to work properly, this 3D model must distinguish between conjugate horizontal ray paths associated with a pair of vertical and transverse modes because of the correspondence with propagation angle orientation and the cross-domain inter-dependencies of the cycle phases.  For this reason, it is important to carefully consider the sign conventions in the convergence factor and the differential topology framework to ensure physical and mathematical consistency.  This model's bi-variate convergence factor is truly an interdependent modal interference calculation utilizing globally-evaluated cycle integrals, so the derivation of the entire convergence factor is presented all at once in an organized manner to hopefully provide a clear and complete perspective of its function.

The WKB modes, \cref{eq:vertWkbModefunctions,eq:transWkbModefunctions}, are substituted into the cross-product summation terms, \cref{eq:crossProductAsFourTerms,eq:crossProductDoublyIncoherentTerm,eq:crossProductVerticalCoherentTerm,eq:crossProductTransverseCoherentTerm,eq:crossProductDoublyCoherentTerm}.  Considering how the pressure-squared varies at a fixed receiver position as the modenumbers vary, e.g. $\npp{\CP}{m}$, all factors outside of the cosines are amplitude factors that vary much more slowly than the interference oscillations represented by the cosines.  Phase differences in the cosine interference oscillations will be approximated with Taylor series expansions for small modenumber differences $\delta_m=m-\hat{m}$ and $\delta_n=n-\hat{n}$.  Therefore, the slower varying amplitude factors may be combined in advance by letting $\hat{m}\raro m$ and $\hat{n}\raro n$ to match the incoherent term's amplitude, c.f. \cref{eq:doublyIncoherentIntegralOverModenumbers}.  Doing so allows factoring out the entire convergence factor expression, $\CC$, before any further approximations, \cref{eq:convergenceFactor1}.

\begin{equation}\begin{aligned}\label{eq:convergenceFactor1}
    &\CP = \frac{S^2}{4}\sum_m^{\IM}\sum_n^{\IN_m}\frac{\CC}{k_{xmn}^2k_{ymn0}k_{ymn}k_{zm0}k_{zm}\wtil{\IL}_{ymn}^2\wtil{\IL}_{zm0}\wtil{\IL}_{zm}} \\
    &\begin{aligned}\CC = 1 &+ 8\sum_{\hat{n}=1}^{n-1} \cos\((k_{xmn}-k_{xm\hat{n}})\abs{x}\)\cos(\IPhi_{ymn0})\cos(\IPhi_{ym\hat{n}0})\cos(\IPhi_{ymn})\cos(\IPhi_{ym\hat{n}}) \\
      &\begin{aligned}+ 32\sum_{\hat{m}=1}^{m-1} \cos\((k_{xmn}-k_{x\hat{m}n})\abs{x}\)&\cos(\IPhi_{ymn0})\cos(\IPhi_{y\hat{m}n0})\cos(\IPhi_{ymn})\cos(\IPhi_{y\hat{m}n}) \\
        &\times\cos(\IPhi_{zm0})\cos(\IPhi_{z\hat{m}0})\cos(\IPhi_{zm})\cos(\IPhi_{z\hat{m}}) \end{aligned} \\
      &\begin{aligned}+ 64\sum_{\hat{m}=1}^{m-1}\sum_{\hat{n}=1}^{n-1} \cos\((k_{xmn}-k_{x\hat{m}\hat{n}})\abs{x}\)&\cos(\IPhi_{ymn0})\cos(\IPhi_{y\hat{m}\hat{n}0})\cos(\IPhi_{ymn})\cos(\IPhi_{y\hat{m}\hat{n}}) \\
        &\times\cos(\IPhi_{zm0})\cos(\IPhi_{z\hat{m}0})\cos(\IPhi_{zm})\cos(\IPhi_{z\hat{m}}) \end{aligned}\end{aligned}
\end{aligned}\end{equation}

Products of cosine pairs differing by modenumber are combined using the product-to-sum trigonometric identity, but the fast interference terms with summed arguments are discarded and only slow interference terms with differenced arguments are retained, \cref{eq:convergenceFactor2}.

\begin{equation}\begin{aligned} \label{eq:convergenceFactor2}
    \CC =& \biggl\{1 + 2\sum_{\hat{n}=1}^{n-1} \cos(\IPhi_{ymn0}-\IPhi_{ym\hat{n}0})\cos(\IPhi_{ymn}-\IPhi_{ym\hat{n}})\cos\((k_{xmn}-k_{xm\hat{n}})\abs{x}\) \\
    &+ 2\sum_{\hat{m}=1}^{m-1} \cos(\IPhi_{ymn0}-\IPhi_{y\hat{m}n0})\cos(\IPhi_{ymn}-\IPhi_{y\hat{m}n})\cos(\IPhi_{zm0}-\IPhi_{z\hat{m}0})\cos(\IPhi_{zm}-\IPhi_{z\hat{m}})\cos\((k_{xmn}-k_{x\hat{m}n})\abs{x}\) \\
    &+ 4\sum_{\hat{m}=1}^{m-1}\sum_{\hat{n}=1}^{n-1} \cos(\IPhi_{ymn0}-\IPhi_{y\hat{m}\hat{n}0})\cos(\IPhi_{ymn}-\IPhi_{y\hat{m}\hat{n}})\cos(\IPhi_{zm0}-\IPhi_{z\hat{m}0})\cos(\IPhi_{zm}-\IPhi_{z\hat{m}})\cos\((k_{xmn}-k_{x\hat{m}\hat{n}})\abs{x}\) \biggr\}
\end{aligned}\end{equation}

Then the product-to-sum formula is used again without dropping terms to combine the remaining cosine products.  At this point it is important to keep the sign convention consistent because of cross-domain interactions, and the sign of the source cycle phase offsets will be associated with conjugate waves initially departing from the source in opposite directions.  The $\CP_{IC}$ term contains only three phase difference cosines, but the $\CP_{CI}$ and $\CP_{CC}$ terms each have five phase difference cosines.  The desired behavior is for source/receiver phase differences to be co-oriented and for vertical/transverse phase differences to be co-oriented, giving $2\times2$ categories of phase differences to co-orient.  Since the cosine function is even and the $k_x\abs{x}$ differences only represent forward propagation from source to receiver, the quintuple cosine product-to-sum identity can be written as combinations of four sign choices ($\sig_1\cdots\sig_4$) and assigned to the four types of phase differences, \cref{eq:convergenceFactor3}.

\begin{equation}\begin{aligned} \label{eq:convergenceFactor3}
    \CC = \sum_{\sigma\in\{-1,+1\}^{\otimes4}}^{\abs{\{\sig\}}=2^4}\biggl\{\frac{1}{16} &+ \frac{1}{8}\sum_{\hat{n}=1}^{n-1}\cos\bigl[\((k_{xmn}-k_{xm\hat{n}})\abs{x}\) + \sigma_1(\IPhi_{ymn0}-\IPhi_{ym\hat{n}0}) + \sigma_2(\IPhi_{ymn}-\IPhi_{ym\hat{n}})\bigr] \\
    &\begin{aligned}+ \frac{1}{8}\sum_{\hat{m}=1}^{m-1}\cos\bigl[((k_{xmn}-k_{x\hat{m}n})\abs{x}) &+ \sigma_1(\IPhi_{ymn0}-\IPhi_{y\hat{m}n0}) + \sigma_2(\IPhi_{ymn}-\IPhi_{y\hat{m}n}) \\
      &+ \sigma_3(\IPhi_{zm0}-\IPhi_{z\hat{m}0}) + \sigma_4(\IPhi_{zm}-\IPhi_{z\hat{m}})\bigr] \end{aligned} \\
    &\begin{aligned}+ \frac{1}{4}\sum_{\hat{m}=1}^{m-1}\sum_{\hat{n}=1}^{n-1}\cos\bigl[((k_{xmn}-k_{x\hat{m}\hat{n}})\abs{x}) &+ \sigma_1(\IPhi_{ymn0}-\IPhi_{y\hat{m}\hat{n}0}) + \sigma_2(\IPhi_{ymn}-\IPhi_{y\hat{m}\hat{n}}) \\
      &+ \sigma_3(\IPhi_{zm0}-\IPhi_{z\hat{m}0}) + \sigma_4(\IPhi_{zm}-\IPhi_{z\hat{m}})\bigr] \biggr\} \end{aligned}
\end{aligned}\end{equation}

The phase differences are expanded by Taylor series to first-order in modenumber differences, $\bigo{\del_m=m-\hat{m}}$ and $\bigo{\del_n=n-\hat{n}}$, about the incoherent solution, $\del_m=0$ and $\del_n=0$.  These Taylor series approximations make use of two derivatives: the total derivative of the open phase integrals with respect to the modenumbers, \cref{eq:openClosedPhaseIntegralJacobian}; and the derivative of the transverse modal eigenvalue $k_x$ with respect to the modenumbers, \cref{eq:dkxdmn}.

\begin{align}
    \Let \delta_m &= m-\hat{m} \in \(0,\del_m^\max\ll\IM\] \tand \delta_n = n-\hat{n} \in \(0,\del_n^\max\ll\IN_m\] \label{eq:modenumberDifferences} \\
    k_{xmn} - k_{x\hat{m}\hat{n}} &\approx \frac{-\pi}{k_x}\[\frac{\wtil{\IL}_{yz}}{\wtil{\IL}_y}\delta_m + \frac{1}{\wtil{\IL}_y}\delta_n\] \label{eq:kxDifference} \\
    \IPhi_{ymn} - \IPhi_{y\hat{m}\hat{n}} &\approx +\pi\[\(\IChi_y\wtil{\IL}_{yz}-\IL_{yz}\)\delta_m + \IChi_y\delta_n\] \label{eq:transModalPhaseDifference} \\
    \IPhi_{zm} - \IPhi_{z\hat{m}} &\approx +\pi\IChi_z\delta_m \label{eq:vertModalPhaseDifference}
\end{align}

The summations are now reordered over modenumber differences $\delta_m$ and $\delta_n$, and maximum mode number differences are set for the upper limits of the summations, $\del_m^\max$ and $\del_n^\max$, see \cref{eq:dirichletKernelConvergenceFactor}.  New convenience notation is introduced here, $\Xi_m$ and $\Xi_n$, which represent the convergence of vertical and transverse cycle phases respectively, \cref{eq:verticalCyclePhaseConvergence} and \cref{eq:transverseCyclePhaseConvergence}.


\begin{align}
    \Xi_n &= \pi\(\abs{\frac{x}{k_x\wtil{\IL}_y}} - \sig_1\IChi_{y0} - \sig_2\IChi_y\) \label{eq:transverseCyclePhaseConvergence} \\
    \Xi_m &= \pi\(\abs{\frac{x\wtil{\IL}_{yz}}{k_x\wtil{\IL}_y}} - \sig_1\[\IChi_{y0}\wtil{\IL}_{yz0}-\IL_{yz0}\] - \sig_2\[\IChi_y\wtil{\IL}_{yz}-\IL_{yz}\] - \sig_3\IChi_{z0} - \sig_4\IChi_z\) \label{eq:verticalCyclePhaseConvergence} \\
    \CC &= \frac{1}{16}\sum_{\sigma\in\{-1,+1\}^{\otimes4}}^{\abs{\{\sig\}}=2^4}\Biggl\{1 + 2\sum_{\hat{m}=1}^{m-1}\cos\(\Xi_m\del_m\) + 2\sum_{\hat{n}=1}^{n-1}\cos\(\Xi_n\del_n\) + 4\sum_{n=1}^{m-1}\sum_{\hat{n}=1}^{n-1}\cos\(\Xi_m\del_m + \Xi_n\del_n\) \Biggr\} \label{eq:dirichletKernelConvergenceFactor}
\end{align}

A multivariate version of Harrison's periodic Gaussian approximation \supercite{harrison2013ray} can be used to simplify the computation and average out the Dirichlet kernel's minor oscillations while still approximately matching the integrated area under the primary peaks, \cref{eq:unsignedPeriodicGaussianConvergenceFactor}.

\begin{gather}
    \Let \IF_m = 1+2\del_m^\max \tand \IF_n = 1+2\del_n^\max \label{eq:periodicGaussianFocusingFactors} \\
    \CC = \frac{\IF_m\IF_n}{16}\sum_{\sigma\in\{-1,+1\}^{\otimes4}}^{\abs{\{\sig\}}=2^4} \exp\[\frac{\IF_m^2}{2\pi}\(\cos\Xi_m-1\)+\frac{\IF_n^2}{2\pi}\(\cos\Xi_n-1\)\] \label{eq:unsignedPeriodicGaussianConvergenceFactor}
\end{gather}

Waves complete the necessary vertical and transverse cycles to arrive at a receiver when the argument of the convergence factor's exponential function becomes zero.  By considering the trajectories of rays leaving the source and using the definitions of the oriented open spectral cycle distances, \cref{eq:vertOpenSpectralCycDist,eq:transOpenSpectralCycDist}, it is possible to associate the plus and minus signs on the source cycle phase terms with specific orientations of waves leaving the source position.  The source sigmas ($\sig_1$ and $\sig_2$) should be substituted with $-1$ to match a wave's initial orientation and to determine how far in the $x$-coordinate an oriented wave must travel before it reaches its first turning point.  Thus the convergence factor is modified to depend implicitly on the propagating wave's initial orientation at the source, and the summation operation is relabeled over just the receiver term sigmas, \cref{eq:signedPeriodicGaussianConvergenceFactor}.  This also effectively accounts for dividing the acoustic power of the energy-flux integrand by two when the angular integration domain is unfolded, and thus a factor of $1/4$ may be extracted from the convergence factor and absorbed by the constant factor in front of the energy flux integral.

\begin{align}
  \Xi_n &= \pi\(\frac{1}{k_x\wtil{\IL}_y}\abs{x} + \IChi_{y0} - \sig_1\IChi_y\) \label{eq:asymmetricTransverseCyclePhase} \\
    \Xi_m &= \pi\(\frac{\abs{\wtil{\IL}_{yz}}}{k_x\abs{\wtil{\IL}_y}}\abs{x} + \[\IChi_{y0}\wtil{\IL}_{yz0}-\IL_{yz0}\] + \IChi_{z0} - \sig_1\[\IChi_y\wtil{\IL}_{yz}-\IL_{yz}\]  - \sig_2\IChi_z\) \label{eq:asymmetricVerticalCyclePhase} \\
    \CC &= \frac{\IF_m\IF_n}{4}\sum_{\sigma\in\{-1,+1\}^{\otimes2}}^{\abs{\{\sig\}}=2^2}\exp\[\frac{\IF_m^2\(\cos\Xi_m-1\)+\IF_n^2\(\cos\Xi_n-1\)}{2\pi}\] \label{eq:signedPeriodicGaussianConvergenceFactor} \\
    \CP &= \frac{S^2}{16\pi^2}\int_{\Omega_\tht}\int_{\Omega_\phi}\Psi\CC\dtht_0\dphi_0 \label{eq:energyFluxIntegralMatchingAsymmetricConvergenceFactor}
\end{align}

\subsection{Adiabatic cycles and the cycle phasor diagram} \label{ssec:adiabaticCyclesAndCyclePhasor}

\begin{figure}[!htb]
  \includegraphics[width=\textwidth]{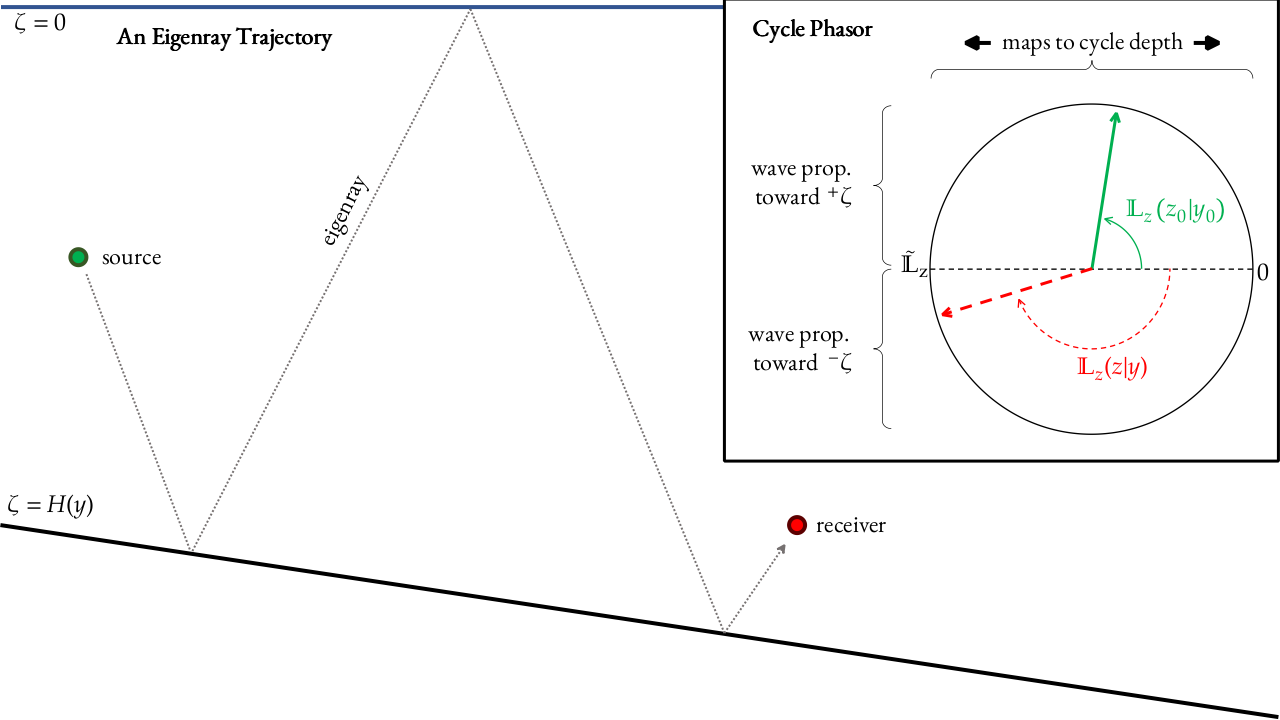}
  \caption{Simplified illustration of a specific eigenray's vertical trajectory and a corresponding cycle phasor diagram.  Partial/open spectral cycle distance integrals are defined from the lesser turning point of a cycle and retain the sign of the wavenumber component.  The entire ray cycle maps bijectively onto the cycle phasor. Thus the positions and propagation directions of a wave at the source or receiver may be distinguished.}
  \label{fig:EigenrayTrajectoryAndCyclePhasor}
\end{figure}

For a more thorough discussion of adiabatic cycle trajectories, the oriented open cycle integrals, and the cycle phasor diagram, the reader may refer to Langhirt's prior investigative work and discussions\supercite{langhirt2025foundations}.  The partial cycle distances have an intuitive physical meaning; they determine how far a ray must travel in order for it to arrive at the next turning point of its cycle.  The convergence factor's partial cycle distance terms represent possible adiabatic eigenray cycle phases that will arrive at a particular receiver.

Because the partial cycle integrals $\IPhi_{zm}(z|y)$, $\IL_{zm}(z|y)$, $\IPhi_{ymn}(y)$, and $\IL_{ymn}(y)$ have been defined in a consistent manner such that they preserve relations between signed wavenumbers and propagation directions, the entire wave cycle can be mapped bijectively onto unique values of the partial cycle integrals.  The waves' trajectories are considered pseudo-periodic since actual periodic ray cycles only exist in a perfectly horizontally-stratified waveguide. However, the adiabatic approximation allows for smoothly transforming range-independent cycles of a local idealized reference waveguide as the actual waveguide varies with range\supercite{brekhovskikh2003fundamentals}.  Wherever the wave is in an adiabatically range-dependent waveguide, it may be mapped onto its pseudo-periodic cycle and the cycle distance can be used as a parameterizing measure of the cycle phase.

Even though the WKB modes and adiabatic approximation allow for general spatial variation of the wavenumber and bathymetry profiles, it is still possible to numerically interpolate the transformations between position within the cycle and the partial spectral cycle distance.  This allows for treating the cycle mapping itself as an invertible operator even if the exact profiles are not integrable as closed-form expressions.  Then any particular wave's trajectory through multiple cycles along the waveguide can be expressed in terms of its vertical and transverse cycle phase functions, \cref{eq:vertCyclePhaseFunction} and \cref{eq:transCyclePhaseFunction}.  \Cref{eq:vertCyclePhaseEvolution} and \cref{eq:transCyclePhaseEvolution} represent the evolution of the vertical and transverse cycle phases as the wave propagates down the waveguide in the forward $x$-direction.  \Cref{eq:vertCyclePhaseOffset} and \cref{eq:transCyclePhaseOffset} are cycle phase offsets that account for the initial source position within the vertical and transverse cycles.

\begin{align}
    \Xi_y(x) &= \Delta\Xi_y + \Xi_{y0} \label{eq:transCyclePhaseFunction} \\
    \Delta\Xi_y(x) &= \tfrac{\pi}{k_x\abs{\wtil{\IL}_y}}\;\abs{x} \label{eq:transCyclePhaseEvolution} \\
    \Xi_{y0}(y_0) &= \pi\IChi_{y0} \label{eq:transCyclePhaseOffset} \\
    \Xi_z(x) &= \Delta\Xi_z + \Xi_{z0} \label{eq:vertCyclePhaseFunction} \\
    \Delta\Xi_z(x) &= \tfrac{\pi\abs{\wtil{\IL}_{yz}}}{k_x\abs{\wtil{\IL}_y}}\abs{x} \label{eq:vertCyclePhaseEvolution} \\
    \Xi_{z0}(y_0,z_0) &= \pi\[\IChi_{y0}\wtil{\IL}_{yz0}-\IL_{yz0}\] + \pi\IChi_{z0} \label{eq:vertCyclePhaseOffset}
\end{align}

In this double mode-sum construction, the entirety of the vertical cycle trajectory depends on where the wave is in its transverse cycle. This is due to the adiabatic approximation holding the modenumber constant as the vertical wavenumber profile and bathymetry vary with range.  The transverse cycle trajectory is much simpler to evaluate, since the vertical modenumber invariance of the adiabatic approximation defines the entire transverse wavenumber profile.

This ability to the map and numerically interpolate the adiabatic cycles allows for two essential techniques to be used.  The first is a transverse mode-stripping kernel, which evaluates the distance until a particular wave exits the computational domain of the waveguide. The second is a cycle-tracking method, which allows tracking individual horizontal trajectories in order to accumulate bottom losses as the waves propagate horizontally along the waveguide.

\subsection{Transverse mode-stripping kernel for the transparent transverse boundary conditions}\label{ssec:transverseModeStrippingKernelTransparentTransverseBoundaryConditions}

The transverse boundaries of the computational domain are transparent, and the model needs a mechanism for stripping modal energy when a wave arrives at and transmits through these boundaries.  To determine when modal energy is stripped from the solution it is necessary to determine when a transverse wave arrives at or is refracted before reaching a transparent boundary of the computational domain.  The mode-stripping kernel was first tested with energy flux methods by calculating the incoherent free-field Green's function solution of a point-source in an unbounded waveguide\supercite{langhirt2025foundations}.

With the typical energy flux assumption that all wavenumber profiles are convex, it is straightforward to check whether a wave is incident or refracted by comparing the horizontal wavenumber at the domain's boundary, $k_{xy}^\min$ or $k_{xy}^\max$, with the transverse eigenvalue, the apexing wavenumber $k_x$.  The distance that a wave must travel before exiting the computational domain can then be determined with conditional logic expressions of transverse cycle distances.  

Any comparative expressions inside of square brackets should be interpreted as \emph{Iverson brackets} which evaluate to $1$ if true and $0$ if false.  The sigma in \cref{eq:phi0positiveConditional}, $\vsig_{y0}$, is true when the wave is initially heading towards the greater turning point.  The rhos in \cref{eq:yminIncidentConditional,eq:ymaxIncidentConditional}, $\rho_y^\min$ and $\rho_y^\max$, are true when the wave is incident on the lesser or greater transverse domain boundary respectively.  The transverse mode-stripping kernel in \cref{eq:transverseModeStrippingKernel}, $\CR_y$, is unity until it switches to zero when a wave has travelled at least the fractional number of cycle distances necessary for the wave to first arrive at a transparent boundary.

\begin{align}
  \vsig_{y0} &= \[\phi_0>0\] \label{eq:phi0positiveConditional} \\
  \rho_y^\min &= \[k_{xy}(y^\min)>k_x\] \label{eq:yminIncidentConditional} \\
  \rho_y^\max &= \[k_{xy}(y^\max)>k_x\] \label{eq:ymaxIncidentConditional} \\
  \CR_y &= \[ [ \rho_y^{\min} \vee \rho_y^{\max} ] \abs{x} \le k_x \( ( \vsig_{y0}\abs{\wtil{\IL}_y} - \IL_{y0} ) + \abs{\wtil{\IL}_y} [ \neg\vsig_{y0}\neg\rho_y^{\min} + \vsig_{y0}\neg\rho_y^{\max} ] \) \] \label{eq:transverseModeStrippingKernel}
\end{align}


For the wedge problem, all of the lesser transverse turning points are refractive.  Each vertical mode's horizontal wavenumber profile vanishes at some point before the wedge apex,  $k_{xy}(y_m^\lwr>0)$, and there is always some transverse turning point $y_{mn}^\lwr > y_{m}^\lwr$ at which the transverse wavenumber profile vanishes $k_y(y_{mn}^\lwr)$, and the wave refracts back towards deeper water.  Similarly, all upper turning points are incident on the greater transverse domain boundary $y^\max$ which is transparent, because $k_{xym}(y)\raro k_{xym}^\max > k_{xmn}$ as $y\raro+\inf$ for all transverse $n$-modes.

\subsection{Accumulated modal attenuation from seafloor interaction along the oriented transverse cycle trajectories} \label{ssec:accumulatedModalAttenuationSeafloorInteractionOrientedTransverseCycleTrajectories}

The bottom loss experienced by a horizontally-refracted vertical mode changes as it propagates forward in the $x$-direction.  This is due to two distinct sources of variation: (a) $y$-dependent bottom sediment properties, and (b) $y$-dependent elevation angle at the seafloor, $\theta(H(y)|y)$, due to adiabatic $\Im$-invariance across a $y$-dependent waveguide depth $H(y)$.  Even with homogeneous sediment properties, the variation due to (b) is unavoidable in a range-dependent environment, so a calculation of the bottom attenuation must account for it in some manner.

Assuming the ray-mode analogy is valid and a waveguide is range-independent, the modal intensity attenuation factor would be evaluated by applying a plane-wave reflection coefficient gradually over the vertical-mode's horizontal cycle distance, as in \cref{eq:rangeIndependentModalIntensityAttenuationKernel}\supercite{tindle1979equivalence,tindle1980connection,tindle1980cycle}

\begin{equation} \label{eq:rangeIndependentModalIntensityAttenuationKernel}
  \CR = \abs{\mathfrak{R}_z}^{\tfrac{2\sqrt{x^2+y^2}}{\ID_z}}
\end{equation}

For a range-dependent waveguide, this expression turns into a product integral along the mode's horizontal trajectory, \cref{eq:rangeDependentIntensityAttenuationKernel}.  In the following equations, $\xi$ is a dummy variable for the $x$-direction which parameterizes the transverse cycles, and the integrals are evaluated from the source at $\xi=0$ to the receiver position at $\xi=x$.  Horizontal distances in the $xy$-plane are evaluated as continuous geometric transformations of transverse cycle distances and wavenumber components.

\begin{equation}\begin{aligned} \label{eq:rangeDependentIntensityAttenuationKernel}
    \CR &= \prod_0^x \abs{\mathfrak{R}_z(\xi)}^{\(\tfrac{\sec\phi(\xi)}{k_{xy}(\xi)\wtil{\IL}_z(\xi)}\)\dxi} \\
    &= \exp\[\int_0^x\frac{\ln\abs{\mathfrak{R}_z(\xi)}}{k_x\wtil{\IL}_z\(y(\xi)\)}\dxi\] \end{aligned}
\end{equation}

The plane-wave reflection-coefficient is dependent on grazing angle at the seafloor, and since the evanescent domain of the vertical modes is not a part of the solution, then only real values of $\tht_{\text{bot}}$ need to be considered.  When $\tht_{\text{bot}}$ is imaginary, its real-part is zero and the reflection-coefficient-magnitude for a perfectly horizontal grazing angle is unity. thus a real vertical mode that refracts before reaching the bottom experiences no attenuation.

\begin{gather}\begin{aligned}\label{eq:6:rangeDependentPlaneWaveReflectionCoefficient}
    \abs{\mathfrak{R}_z(\xi)} &= \left\{\;\begin{aligned} \abs{\mathfrak{R}_z\(\tht_{\text{bot}}(\xi)\)} &&& \tfor k\(H\(y(\xi)\)|y(\xi)\) > k_{xy}\(y(\xi)\) \\ 1 &&& \tfor k\(H\(y(\xi)\)|y(\xi)\) \le k_{xy}\(y(\xi)\) \end{aligned}\right. \\
    &= \abs{\mathfrak{R}_z\(\Real{\acos\(\frac{k_{xy}\(y(\xi)\)}{k\(H\(y(\xi)\),y(\xi)\)}\)}\)} \\
\end{aligned}\end{gather}

The only remaining unknown is the transverse coordinate of the vertical mode's horizontal trajectory, $y(\xi)$, which can be determined using cycle-tracking.  For the cycle-tracking method, the transverse cycle phase is interpolated back to a $y$-position in the transverse cycle by numerically inverting the open spectral cycle distance operator, \cref{eq:transOpenSpectralCycDistOperator}.

\begin{equation}\begin{gathered} \label{eq:transOpenSpectralCycDistOperator}
      \SL_y\{\;\}\;:\;\[y^\lwr,y^\upr\]\raro\[0,\abs{\wtil{\IL}_y}\] \\
      \SL_y\{y\} = \abs{\IL_y(y)} \tand \SL_y^{-1}\{\SL_y\{y\}\} = y
\end{gathered}\end{equation}

First the accumulated transverse cycle phase at some forward $x$-distance is evaluated by applying the transverse open spectral cycle distance operator, \cref{eq:transAccumCyclePhase}.

\begin{equation}
      \Xi_y(x) = \frac{\pi}{k_x\abs{\wtil{\IL}_y}}\(\abs{x}+k_x\IL_{y0}\) \label{eq:transAccumCyclePhase} \\
\end{equation}

Complete cycles of $2\pi$ radians are subtracted off from the accumulated transverse cycle phase, \cref{eq:transAccumCyclePhase}, to obtain the principal transverse cycle phase, \cref{eq:transPrincipalCyclePhase}.  The inverse transverse open spectral cycle distance operator is used to map the principal transverse cycle phase magnitude, \cref{eq:transPrincipalCyclePhaseMagnitude}, to the transverse open spectral cycle distance magnitude ($\abs{\IL_y}(x)$), the transverse wavenumber magnitude ($\abs{k_y(x))}$ in \cref{eq:transverseCycleMomentum}), and the $y(x)$ position in the transverse cycle, \cref{eq:transverseCyclePosition}.  Then the principal transverse cycle phase orientation, \cref{eq:transPrincipalCyclePhaseOrientation}, determines the sign of the transverse wavenumber, \cref{eq:transverseCycleMomentum}.

\begin{align}
  \angle\Xi_y(x) &= \Xi_y(x) \bmod 2\pi \in [0,\pi)\cup[\pi,2\pi) \label{eq:transPrincipalCyclePhase} \\
  \abs{\angle\Xi_y}(x) &= \pi - \abs{\angle\Xi_y(x)-\pi}  \in [0,\pi)\cup[\pi,0) \label{eq:transPrincipalCyclePhaseMagnitude} \\
  \sig_y(x) &= \sgn\(\pi-\angle\Xi_y(x)\) \in \{+1\}\cup\{-1\} \label{eq:transPrincipalCyclePhaseOrientation}
\end{align}

\begin{align}
  y(x) &= \SL_y^{-1}\left\{\frac{\abs{\wtil{\IL}_y}}{\pi}\cdot\abs{\angle\Xi_y}(x)\right\} \label{eq:transverseCyclePosition} \\
  k_y(x) &= \sig_y(x)\cdot \abs{k_y(y(x))} \label{eq:transverseCycleMomentum}
\end{align}

\subsection{Cross-domain cycle-tracking and simulated 3d ray trajectories} \label{ssec:crossDomainCycleTrackingSimulated3dRayTrajectories}

A similar construction can be formulated for the vertical cycles as well, but the vertical cycle phase is $y$-dependent, therefore the transverse cycle phase must be evaluated before the vertical cycle phase can be evaluated.  As with the transverse cycles, an operator can be defined that maps vertical position within the vertical cycles to the vertical open spectral cycle distance for a particular choice of $y$-position, and this operator can be numerically interpolated and inverted, \cref{eq:vertOpenSpectralCycDistOperator}.

\begin{equation}\begin{gathered} \label{eq:vertOpenSpectralCycDistOperator}
    \SL_z\{\;|\;\}\;:\;\[z^\lwr(y),z^\upr(y)\]\raro\[0,\abs{\wtil{\IL}_z(y)}\] \\
    \SL_z\{z|y\} = \abs{\IL_z(z|y)} \tand \SL_z^{-1}\{\SL_z\{z|y\}|y\} = z
\end{gathered}\end{equation}

The accumulated vertical cycle phase at some forward $x$-distance is evaluated by applying the vertical open spectral cycle distance operator, \cref{eq:vertAccumCyclePhase}.  This expression now involves evaluating the cross-domain cycle integral, $\wtil{\IL}_{yz}(y)$, which accumulates the number of vertical half-cycles per transverse half-cycle, effectively functioning as a cross-domain cycle phase transformation factor.  The cross-domain cycle phase terms are evaluated at the boundaries of the $x$-interval of propagation, and thus seem to be functioning as a sort of pre-integrated correction to the vertical cycle phase (c.f. the fundamental theorem of calculus).

\begin{equation}
  \Xi_z(x) = \pi\(\frac{\abs{x}\,\abs{\wtil{\IL}_{yz}}}{k_x\abs{\wtil{\IL}_y}}+\abs{\wtil{\IL}_{yz}}\evalover{\frac{\vsig_\ups\abs{\IL_{yz}(\ups)}}{\abs{\wtil{\IL}_{yz}}}-\frac{\vsig_\ups\abs{\IL_y(\ups)}}{\abs{\wtil{\IL}_y}}}{(\ups,\vsig_\ups)=(y(0),\vsig_y(0))}{(\ups,\vsig_\ups)=(y(x),\vsig_y(x))} - \frac{\IL_{z0}}{\abs{\wtil{\IL}_{z0}}}\) \label{eq:vertAccumCyclePhase} \\
\end{equation}

Complete cycles of $2\pi$ radians are subtracted off from the accumulated vertical cycle phase, \cref{eq:vertAccumCyclePhase}, to obtain the principal vertical cycle phase, \cref{eq:vertPrincipalCyclePhase}.  The inverse vertical open spectral cycle distance operator is used to map the principal vertical cycle phase magnitude, \cref{eq:vertPrincipalCyclePhaseMagnitude}, to the vertical open spectral cycle distance magnitude, $\abs{\IL_z}(y(x))$, the vertical wavenumber magnitude, $\abs{k_z(z(x),y(x))}$ in \cref{eq:verticalCycleMomentum}, and the $z(x)$ position in the vertical cycle, \cref{eq:verticalCyclePosition}.  Then the vertical principal cycle phase orientation, \cref{eq:vertPrincipalCyclePhaseOrientation}, determines the sign of the vertical wavenumber, \cref{eq:verticalCycleMomentum}.

\begin{align}
  \angle\Xi_z(x) &= \Xi_z(x) \bmod 2\pi \in [0,\pi)\cup[\pi,2\pi) \label{eq:vertPrincipalCyclePhase} \\
  \abs{\angle\Xi_z}(x) &= \pi - \abs{\angle\Xi_z(x)-\pi}  \in [0,\pi)\cup[\pi,0) \label{eq:vertPrincipalCyclePhaseMagnitude} \\
  \sig_z(x) &= \sgn\(\pi-\angle\Xi_z(x)\) \in \{+1\}\cup\{-1\} \label{eq:vertPrincipalCyclePhaseOrientation}
\end{align}

\begin{align}
  z(x) &= \SL_z^{-1}\left\{\frac{\abs{\wtil{\IL}_z(y(x))}}{\pi}\cdot\abs{\angle\Xi_z}(x)\middle|y(x)\right\} \label{eq:verticalCyclePosition} \\
  k_z(x) &= \sig_z(x)\cdot \bigl|k_z(z(x)|y(x))\bigr| \label{eq:verticalCycleMomentum}
\end{align}

The set of equations for $y(x)$ and $k_y(x)$, \cref{eq:transverseCyclePosition,eq:transverseCycleMomentum}, and $z(x)$ and $k_z(x)$, \cref{eq:verticalCyclePosition,eq:verticalCycleMomentum}, are analogous to the standard system of ray equations for position and momentum of a wave in the geometric acoustic limit\supercite{jensen2011computational}.  However, these trajectories are not expected to match the trajectories of 3D ray tracing methods.  The adiabatic approximation will continuously curve the propagation trajectory in an isovelocity environment with range-dependent bathymetry, whereas 3D ray tracing methods will only alter ray trajectories when the ray encounters a refractive interface or reflective boundary.

\subsection{Directivity kernel for approximating boundary interference effects} \label{ssec:directivityKernelApproximatingBoundaryInterferenceEffects}

When a simple acoustic source is brought near a planar acoustic boundary, the source field superimposes with the reflections to produce an interference pattern commonly referred to as Lloyd's mirror.  In the far-field (small-angle) asymptotic limit, the interference phase shift asymptotically limits the apparent average power radiated to somewhere between that of a perfect bipole and perfect dipole\supercite{brekhovskikh2012acoustics,brekhovskikh2013acoustics,pierce2019acoustics}.  This limiting behavior of the far-field power depends on: (a) the reflection coefficient of the boundary, and (b) the source or receiver's distance from the boundary relative to the acoustic wavelength.  

The energy flux derivation does not inherently incorporate this type of interference, so the model needs an additional mechanism to modify the acoustic power in the far-field asymptotic limit.  This effect can be approximated by including another integration kernel that weights the angular distribution of acoustic intensity with source and receiver directivity patterns that correspond to the far-field asymptotic limit of a monopole brought within several wavelengths of a physical boundary \supercite{holland2010propagation}.  This is the same type of modification made by the ``semicoherent'' run-type option of the ray tracing model ``Bellhop'' developed by Mike Porter\supercite{porter2011bellhop}.

\Cref{eq:directivityWavenumber} expresses a range-dependent wavenumber at the surface and bottom, \cref{eq:directivityGrazingAngles} gives the range-dependent grazing angles at the surface and bottom, and \cref{eq:directivityReflectionCoefficients} defines the range-dependent plane-wave reflection coefficient at the surface and bottom.  The directivity pattern for far-field intensity of a monopole near a single boundary in an isovelocity range-independent medium is given in \cref{eq:lloydMirrorDirectivityPattern}, but is modified to map the reflection coefficient's grazing angle to the elevation propagation angle at the monopole.  The full directivity kernel, \cref{eq:lloydMirrorDirectivityKernel}, is the product of four directivity patterns corresponding to combinations of source/receiver and surface/bottom reflections.

\begin{align}
  k^{\{\min,\max\}}(\ups) &= k(z^{\{\min,\max\}}(\ups)|\ups) \twhere z^{\{\min,\max\}}(\ups) = \{0,H(\ups)\} \label{eq:directivityWavenumber} \\
  \tht_\angle^{\{\min,\max\}}(\ups) &= \tht(z^{\{\min,\max\}}(\ups)|\ups) = \Real{\acos\(\frac{k_{xy}(\ups)}{k^{\{\min,\max\}}(\ups)}\)} \label{eq:directivityGrazingAngles} \\
  \mathfrak{R}_z^{\{\min,\max\}}(\ups) &= \mathfrak{R}_z^{\{\min,\max\}}\(\tht_\angle^{\{\min,\max\}}(\ups)|\ups\) \label{eq:directivityReflectionCoefficients} \\
  \CD^{\{\min,\max\}}\(\zeta|\ups\) &= \half\abs{1+\mathfrak{R}_z^{\{\min,\max\}}\(\ups\)\exp\(i2k_z(\zeta|\ups)\abs{z^{\{\bot,\top\}}-\zeta}\)}^2 \label{eq:lloydMirrorDirectivityPattern} \\
  \CD &= \CD^\min\cdot\CD^\max\cdot\CD_0^\min\cdot\CD_0^\max \twhere (\zet|\ups) = \{(z,y),(z_0,y_0)\} \label{eq:lloydMirrorDirectivityKernel}
\end{align}


\subsection{3D semi-coherent solid-angle energy flux model for a transversely range-dependent lossy waveguide} \label{ssec:threeDimensionalSemiCoherentSolidAngleEnergyFluxModelTransverselyRangeDependentLossyWaveguide}

Inserting all of the integration kernels and ancillary cycle calculations into the source-oriented doubly-incoherent energy flux integral, \cref{eq:wkbLimiterModification}, yields the final 3D solid-angle energy flux model, \cref{eq:3DSAEF}.  This model integrates the acoustic intensity over differential solid-angles and includes distinct integration kernels to account for: (a) modal energy scaling with WKB amplitudes, $\Psi$ from \cref{eq:modalAmplitudeSquaredKernel}; (b) focusing of modal energy around source-receiver eigenrays to modulate the $x$-dependent field structure, $\mathcal{C}$ from \cref{eq:signedPeriodicGaussianConvergenceFactor}; (c) transverse-mode-stripping for waves incident on transparent transverse boundaries using extinction distances calculated from oriented source azimuthal angles, $\mathcal{R}_y$ from \cref{eq:transverseModeStrippingKernel}; (d) bottom attenuation as a range-dependent product-integral along a vertical mode's equivalent oriented horizontal ray trajectory calculated from transverse-cycle-tracking, $\mathcal{R}_z$ from \cref{eq:rangeDependentIntensityAttenuationKernel}; and (e) source and receiver directivity patterns for approximating Lloyd's mirror effects when a monopole is brought within several wavelengths of a physical boundary, $\mathcal{D}$ from \cref{eq:lloydMirrorDirectivityKernel}.

\begin{gather}\begin{aligned}\label{eq:3DSAEF}
    \CP &= \int_{\Omega_\theta}\int_{\Omega_\phi}\Psi\;\CD\;\CR_y\;\CR_z\;\CC\;\dphi_0\dtht_0 \\
    \TL &= -10\log_{10}\(\frac{\CZ_0}{\CZ}\;\CP\)
\end{aligned}\end{gather}

\section{IMPLEMENTATION, DEMONSTRATION, AND COMPARISON} \label{sec:implementationDemonstrationComparison}

\subsection{Outline of the calculation's implementation} \label{ssec:outlineCalculationImplementation}

This 3D solid-angle energy flux model was implemented in Matlab for computing the transmission loss (TL) field in environments with an isovelocity lossless watercolumn, range-independent sediment properties, and a $y$-range-dependent bathymetry profile; i.e. a wedge or trough of constant cross-section.  The model implementation is a vectorized calculation using multi-dimensional arrays.  The main calculation makes use of several distinct numerical grids: the environmental transverse $y$-grid, the receiver position $(x,y,z)$-grid, the source angular $(\theta_0,\phi_0)$-grid, and the interpolation grids used to sample the cycle calculations over queried $x$, $y$, and $\theta$ values.

The model first needs to sample environmental precalculations such as wavenumber profiles and interpolations of cycle integrals, and these precalculations could be made exclusively dependent on only the environmental parameters and not the source/receiver parameters.  Future versions of the model could execute the environmental precalculation routine separately from the field calculation routine, and have the precalculations stored on disk as lookup tables for multiple field calculations using the same environment.  The adiabatic invariance mapping is accomplished using Matlab's builtin contour search routines to find iso-vertical-modenumber horizontal-wavenumber profiles, $k_{xym}(y|k_{xym0})$ such that $\Im(y_0)=\Im(y)$.  

Then all of the wavenumber profiles and open(partial)/closed(total) cycle integrals are sampled and interpolated, all interpolations are constructed as piecewise-linear profiles, and the cycle integrals are evaluated exactly for piecewise-linear profiles from one turning point to the other.  Then the integration kernels are evaluated for each receiver position and source propagation angle.  Lastly, the energy flux integral over source propagation angles is evaluated using a composite Simpson's rule quadrature scheme.  The calculation steps that are the most computationally time-consuming are those with the highest parametric degrees of freedom, sampling the open cycle integral interpolations and evaluating the bivariate convergence factor, so these calculations were parallelized to speed up the computation.

The 3D solid-angle energy flux model implementation that is used in this paper is licensed as open-source software under the MIT License and is hosted on Github: 
\url{https://github.com/marklanghirt/Tethys}\supercite{langhirt2025tethys}.

\subsection{The canonical ASA wedge test environment and inter-model transmission loss comparisons} \label{ssec:canonicalAsaWedgeTestEnvironmentTransmissionLossComparisons}

\begin{figure}[!htb]
  \includegraphics[width=\textwidth]{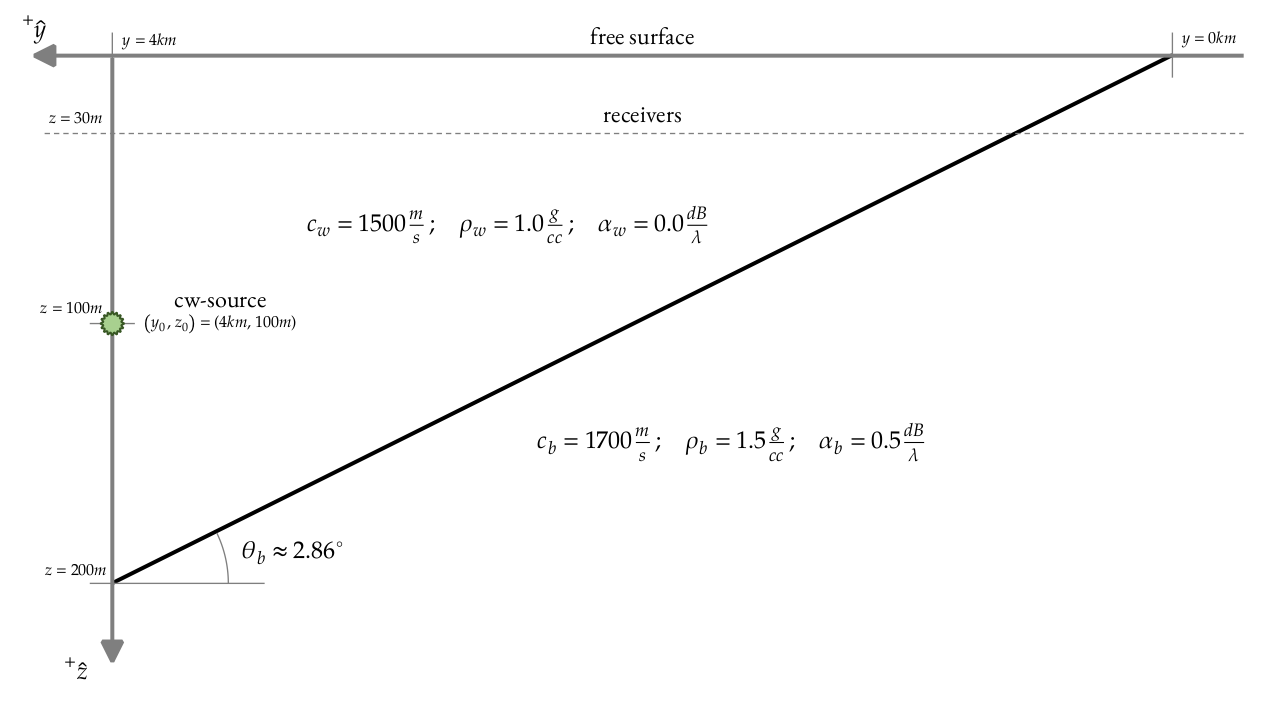}
  \caption{Replication of Figure 1 from Jensen and Ferla (1990)\supercite{jensen1990numerical}, showing the standard wedge geometry for testing 3D ocean acoustic propagation models.}
  \label{fig:asaWedgeGeometry}
\end{figure}

The 3D solid-angle energy flux model's TL field is compared against other 3D ocean acoustic propagation models' TL fields for the canonical ASA wedge problems I and III as described in Jensen and Ferla's 1990 paper\supercite{jensen1990numerical}, a diagram of which is reproduced here in \cref{fig:asaWedgeGeometry}.  

Wedge problem I (ideal) in \cref{ssec:wej1} is an isovelocity, lossless, and impenetrable wedge environment ($c_w=1500\unit{\tfrac{m}{s}}$, $\rho_w=1\unit{\tfrac{g}{cc}}$, and $\alf_w=0\unit{\tfrac{dB}{\lam}}$) with a rigid bottom boundary and slope of approximately $2.86\deg$ from the horizontal.  An analytic integral transform solution is available for the ideal wedge that was first derived by Mike Buckingham\supercite{buckingham1984acoustica,buckingham1984acousticb} and can also be found in George Frisk's book, Ocean Acoustics\supercite{frisk1994ocean}.

\begin{gather}
  \alf_\nu=\(\nu-\tfrac{1}{2}\)\tfrac{\pi}{\bet} \tand \kap=\sqrt{k^2-k_x^2} \label{eq:analyticIdealWedgeEigenvalues} \\
  r_< = \opmin\(r_0,r\) \tand r_> = \opmax\(r_0,r\) \label{eq:analyticIdealWedgeRange} \\
  p(r,\tht,x) = \frac{i2\pi}{\bet}\sum_\nu\sin\alf_\nu\tht_0\sin\alf_\nu\tht\int_{\ngtv\inf}^{\pstv\inf}J_{\alf_\nu}(\kap r_<)H_{\alf_\nu}^{(1)}(\kap r_>)e^{ik_xx}\dk_x \label{eq:analyticIdealWedgePressure}
\end{gather}

One of the primary characteristics observed in the analytic ideal wedge solution, \crefrange{eq:analyticIdealWedgeEigenvalues}{eq:analyticIdealWedgePressure}, are the existence of so-called wedge modes which correspond with the angular modenumbers ($\nu$).  These wedge modes are most easily distinguished at lower frequencies when the acoustic intensity lobes are larger and fewer in number.  For the $25\unit{Hz}$ TL field solutions depicted in \cref{fig:wej1f025y4000,fig:wej1f025z30}, the locations of acoustic intensity lobes are traced with dashed black lines for ease of direct comparison.

Wedge problem III in \cref{ssec:wej3} is a lossy penetrable wedge environment with the same boundary geometry and isovelocity watercolumn properties, but with an attenuative acoustic sediment ($c_b=1700\unit{\tfrac{m}{s}}$, $\rho_b=1.5\unit{\tfrac{g}{cc}}$, $\alf_b=0.5\unit{\tfrac{dB}{\lam}}$).  There is no closed-form integral transform solution available for the lossy, penetrable wedge environment.  The 3D split-step Fourier wide-angle parabolic equation model has been used extensively for low-frequency 3D ocean acoustic propagation modeling, and is used here to provide a trusted ground truth for accurate comparisons of the TL field in the lossy penetrable wedge environment.

Then lastly in \cref{ssec:quadRays}, 3D adiabatic cycle trajectories in an ideal wedge environment are calculated with the model's cycle tracking method and are plotted for demonstration purposes.  These adiabatic cycle trajectories are not compared directly with 3D ray tracing trajectories since they should inherently produce distinctly different trajectories.  For 3D ray tracing, the ray trajectory headings would only change when encountering a reflective or refractive boundary, which would only be the boundaries of the waveguide in an isovelocity environment.  For the 3D energy flux model, the adiabatic vertical modes approximation for transverse range-dependence produces an effective horizontal wavenumber profile, and therefore the adiabatic cycle trajectories should compare better with horizontal ray trajectories from a vertical modes horizontal rays solution.

All of the energy flux TL solutions provided in this section were calculated with inclusion of the WKB limiters from \cref{eq:wkbLimiterModification}, the Lloyd's mirror directivity kernel provided in \cref{eq:lloydMirrorDirectivityKernel}, and the finite-frequency elevation angle floor described for \cref{eq:customSourceThetaDomain}.  The finite-frequency elevation angle floor has also been applied to the ray-tracing solutions since ray-tracing models also make use of high-frequency approximations and a continuum of propagation angles.  The finite-frequency approximation used in the high-frequency continuum models reveals a horizontal shadow zone that corresponds to the exterior of the lowest wedge mode's envelope and makes the horizontal refraction of acoustic intensity more apparent.

The purpose of including the Lloyd's mirror directivity kernel in the energy flux model is to reintroduce vertical modal interference due to the boundary reflections so that the wedge modes can be identified and horizontal refraction can be clearly demonstrated.  However, there is physical justification for approximating vertical modal interference from surface and bottom boundaries using the Lloyd's Mirror directivity patterns.  The Lloyd's mirror directivity pattern was intended to be effectively the same as that used by Bellhop3D's semi-coherent solution, though there does appear to be some difference due to the lack of clearly defined wedge modes in the semi-coherent ray tracing solutions.  Without the directivity pattern, the energy flux solution would appear more like a smoothed gradient in an isovelocity environment, though the mechanisms enabling horizontal refraction (the transverse convergence) would still be present and functional.

\subsection{Lossless Impenetrable Wedge I} \label{ssec:wej1}

For all side-by-side comparisons of 2D TL plots in this ideal wedge subsection, solution (a) in the top left corner is from the 3D solid-angle energy flux model with transverse $y$-convergence only, solution (b) in the top-right corner is the semi-coherent ray-tracing solution calculated by Bellhop3D\supercite{porter2011bellhop}, solution (c) in the bottom-left corner is the analytic ideal wedge solution\supercite{buckingham1984acoustica,buckingham1984acousticb,frisk1994ocean}, and solution (d) in the bottom-right corner is the coherent ray-tracing solution calculated by Bellhop3D\supercite{porter2011bellhop}.

\begin{figure}[!htb]
  \includegraphics[width=\textwidth]{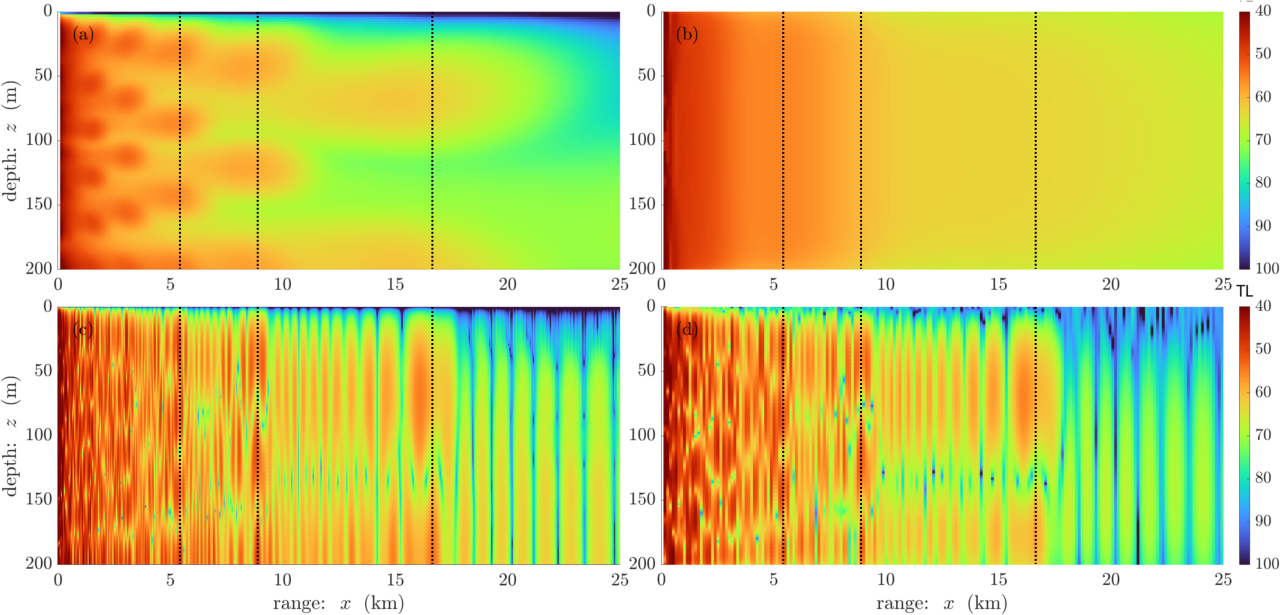}
  \caption{Side-by-side comparison of a 2D TL slice at $y=4\unit{km}$ range for the ideal wedge I at $25\unit{Hz}$ frequency.  The models included are: (a) the 3D solid-angle energy flux model, (b) the semi-coherent 3D ray tracing model, (c) the analytic solution for the ideal wedge, and (d) the coherent 3D ray tracing model.  Some of the wedge mode envelopes are marked by dashed black lines for easier comparisons.}
  \label{fig:wej1f025y4000}

  \includegraphics[width=\textwidth]{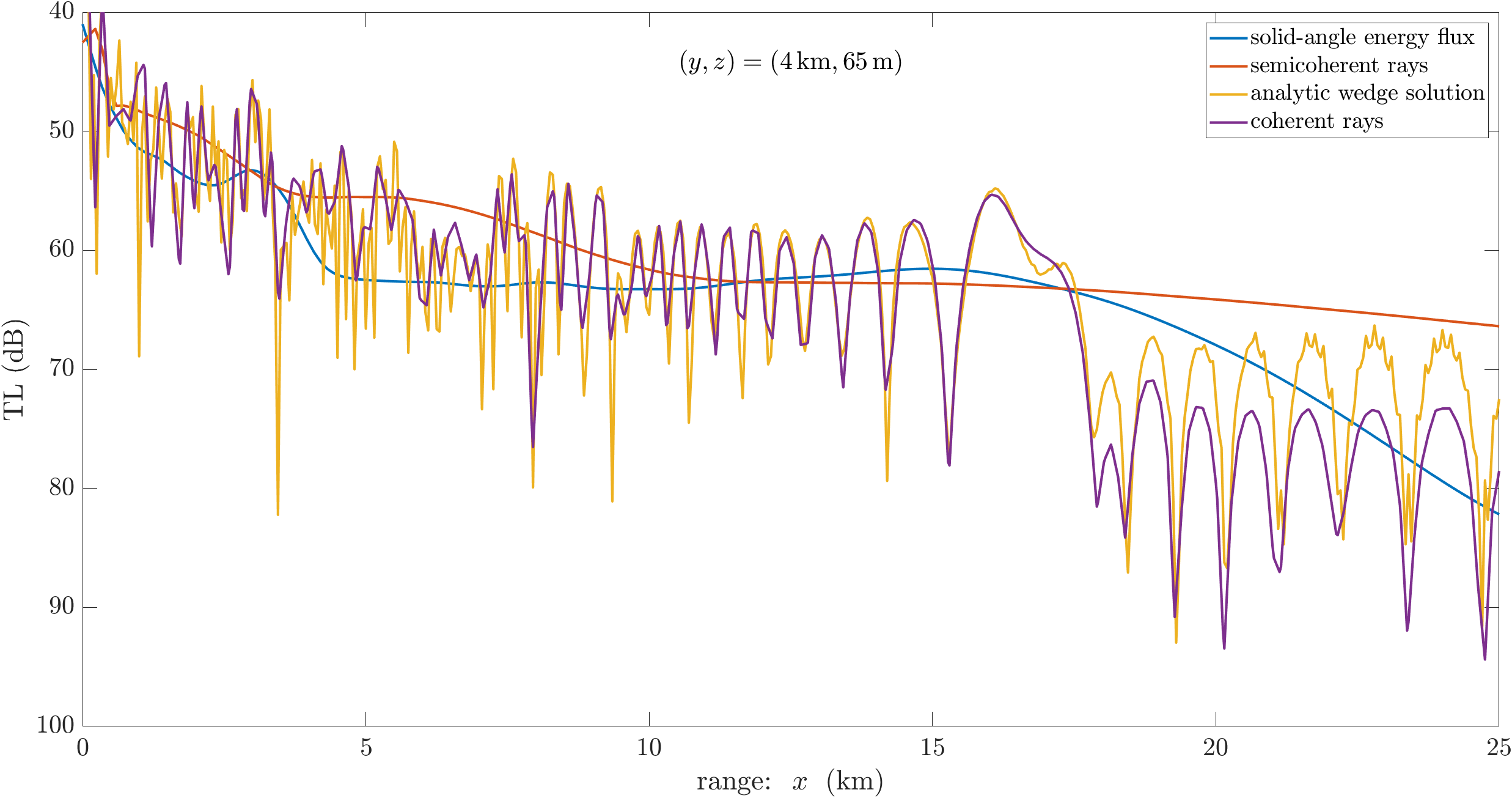}
  \caption{Comparison of TL for the ideal wedge I at $25\unit{Hz}$ frequency along a line array of receivers at $(y,z)=(4\unit{km},65\unit{m})$.}
  \label{fig:wej1f025y4000z65}
\end{figure}

\Cref{fig:wej1f025y4000} shows a 2D TL slice at $4\unit{km}$ transversely from the wedge apex, and the longitudinal positions of wedge mode envelopes from the analytic solution are marked by vertical dashed lines.  There does appear to be corresponding intensity lobes in the energy flux solution that are centered on these same locations, but the extent and shape of these lobes differ due to the absence of fine-scale modal interference.  The maximum transverse mode number difference, $\delta_n^\max$ from the periodic Gaussian focusing factors in \cref{eq:periodicGaussianFocusingFactors}, is another parameter that affects the size and shape of the focused intensity lobes, but this focusing is likely limited to some proportion of the cycle distances since only the large-scale interference of neighboring modes has been retained.

\begin{figure}[!htb]
  \includegraphics[width=\textwidth]{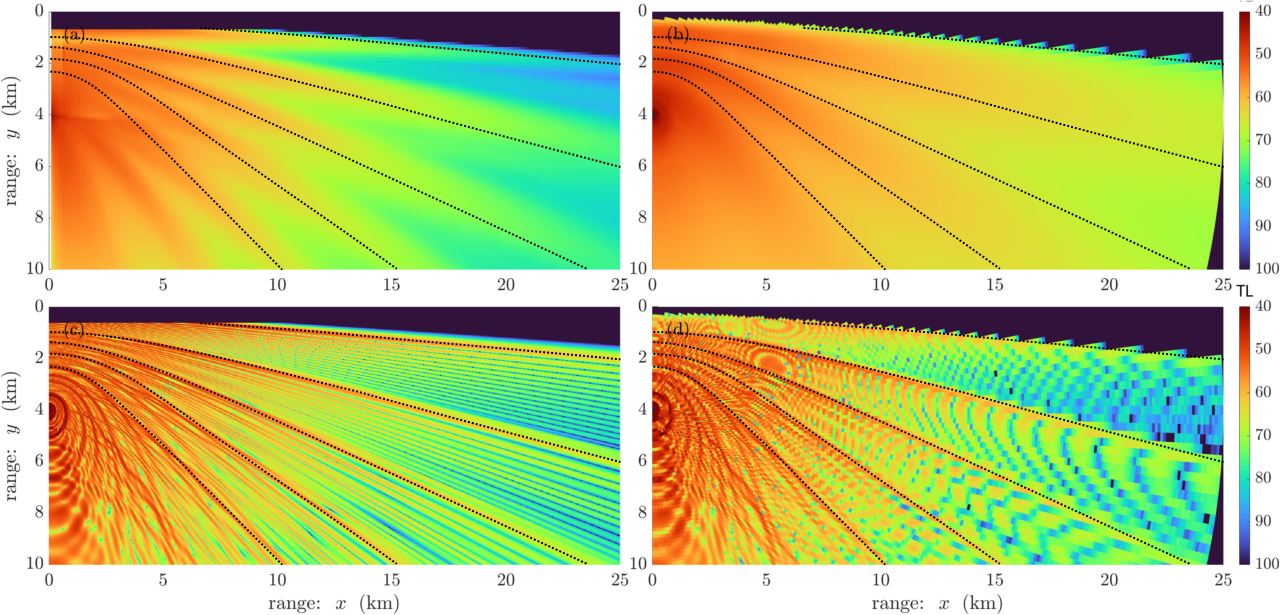}
  \caption{Side-by-side comparison of a 2D TL slice at $z=30\unit{m}$ depth for the ideal wedge I at $25\unit{Hz}$ frequency.  The models included are: (a) the 3D solid-angle energy flux model, (b) the semi-coherent 3D ray tracing model, (c) the analytic solution for the ideal wedge, and (d) the coherent 3D ray tracing model.  Some of the wedge mode envelopes are marked by dashed black curves for easier comparisons.}
  \label{fig:wej1f025z30}

  \includegraphics[width=\textwidth]{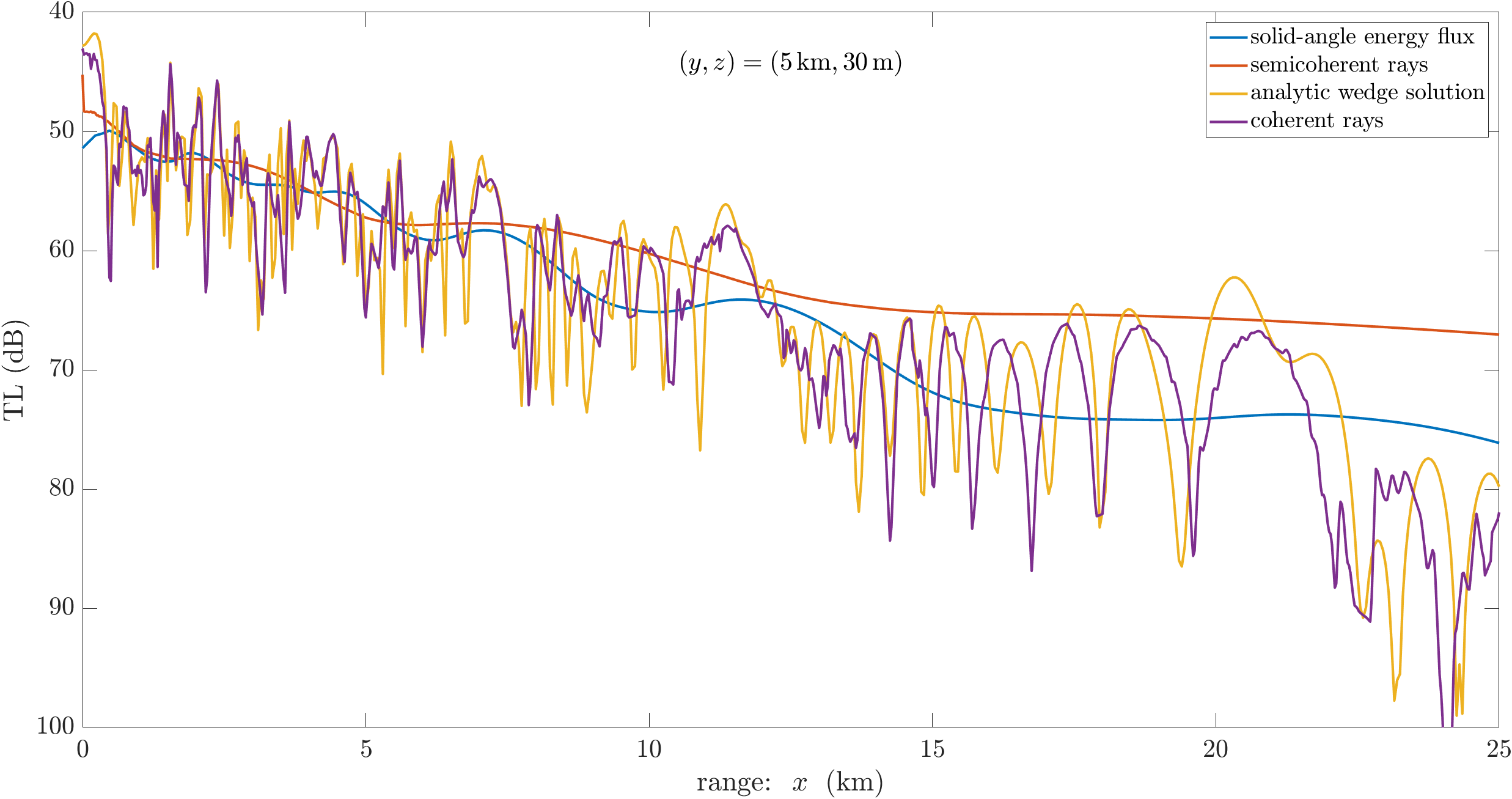}
  \caption{Comparison of TL for the ideal wedge I at $25\unit{Hz}$ frequency along a line array of receivers at $(y,z)=(5\unit{km},30\unit{m})$.}
  \label{fig:wej1f025z30y5000}
\end{figure}

One of the features that is already present in \cref{fig:wej1f025y4000}, but can also be seen in \cref{fig:wej1f025z30} and \cref{fig:wej3f025z30}, is the difference between the semi-coherent ray solution and the energy flux solution in the incoherent background intensity levels at further ranges and low frequency.  However, this difference does not appear to be present in the coherent ray tracing solutions, which seem to align nicely with the overall TL levels and interference phases of the analytic solutions.  This TL offset, approximately $10\unit{dB}$ at $25\unit{km}$ range in \cref{fig:wej1f025y4000z65} and \cref{fig:wej1f025z30y5000}, may be due to the inherent difference in the degree of horizontal refraction between the adiabatic modes approximation and the 3D ray tracing solution.  The effective horizontal wavenumber profile induced by the adiabatic modes approximation has a frequency-dependent rate of horizontal refraction and the wedge modes' horizontal envelopes are known to be narrower at lower frequencies, however the 3D ray tracing solutions only refract horizontally in the isovelocity waveguide when the rays are incident on the sloped bottom.  These are curious features that could be investigated further with higher resolution model executions on more capable computers.

Even though the energy flux model makes use of several high-frequency continuum approximations, it is pleasantly surprising to see such good agreement in the $25\unit{Hz}$ solutions.  One nice feature to see in \cref{fig:wej1f025z30} is the wedge mode envelopes, traced with dashed black lines, agreeing in shape and position with those in the analytic solution.  A combination of two physical mechanisms in the energy flux solution is inducing the simulation of these wedge modes.  Firstly, the vertical Lloyd's mirror directivity patterns generate an interference pattern between the source and each receiver location.  Then the transverse convergence factor is focusing these beams onto trajectories that are refracted at the rate determined by the adiabatic approximation's effective horizontal wavenumber profile and resulting transverse wave cycle.  

\begin{figure}[!htb]
  \includegraphics[width=\textwidth]{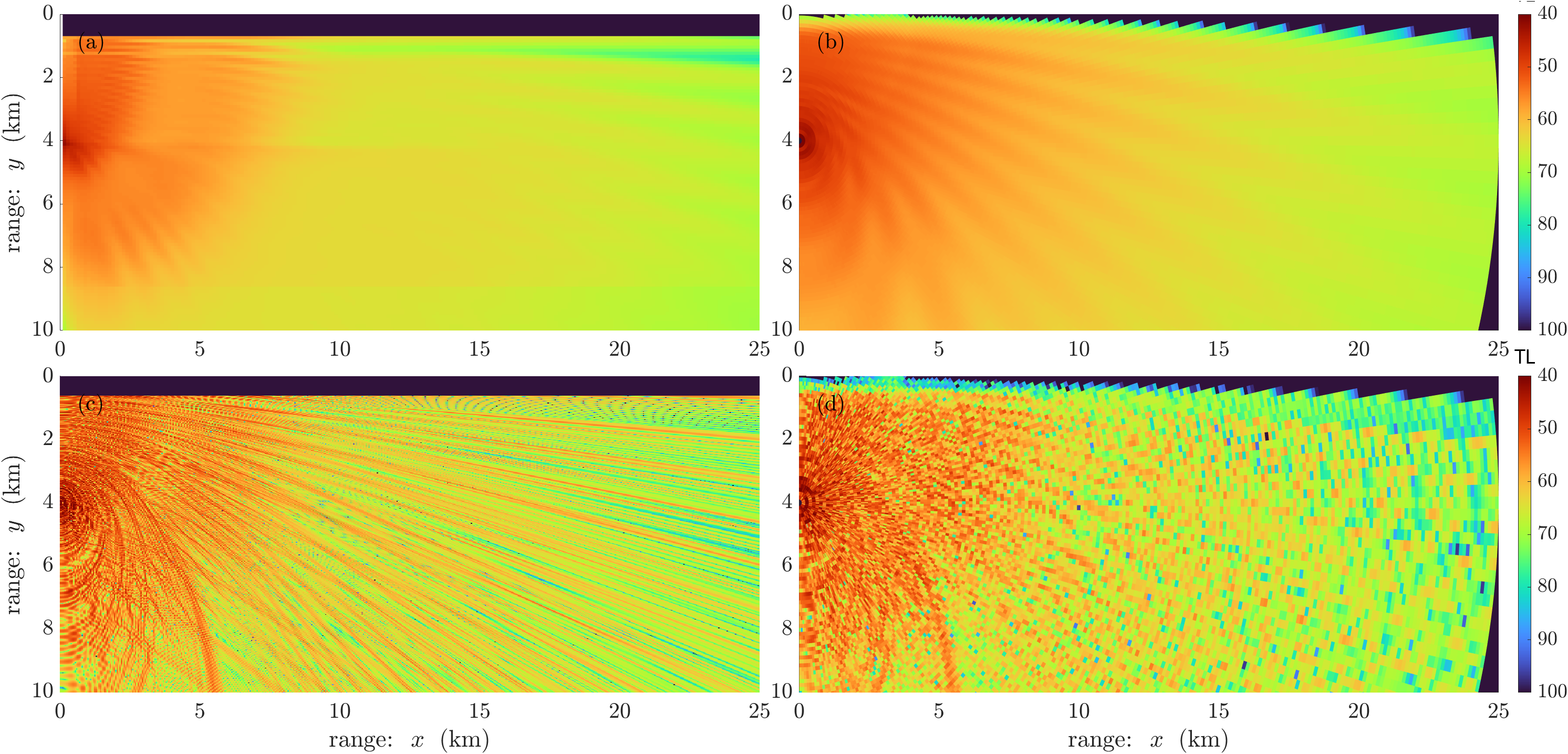}
  \caption{Side-by-side comparison of a 2D TL slice at $z=30\unit{m}$ depth for the ideal wedge I at $75\unit{Hz}$ frequency.  The models included are: (a) the 3D solid-angle energy flux model, (b) the semi-coherent 3D ray tracing model, (c) the analytic solution for the ideal wedge, and (d) the coherent 3D ray tracing model.}
  \label{fig:wej1f075z30}

  \includegraphics[width=\textwidth]{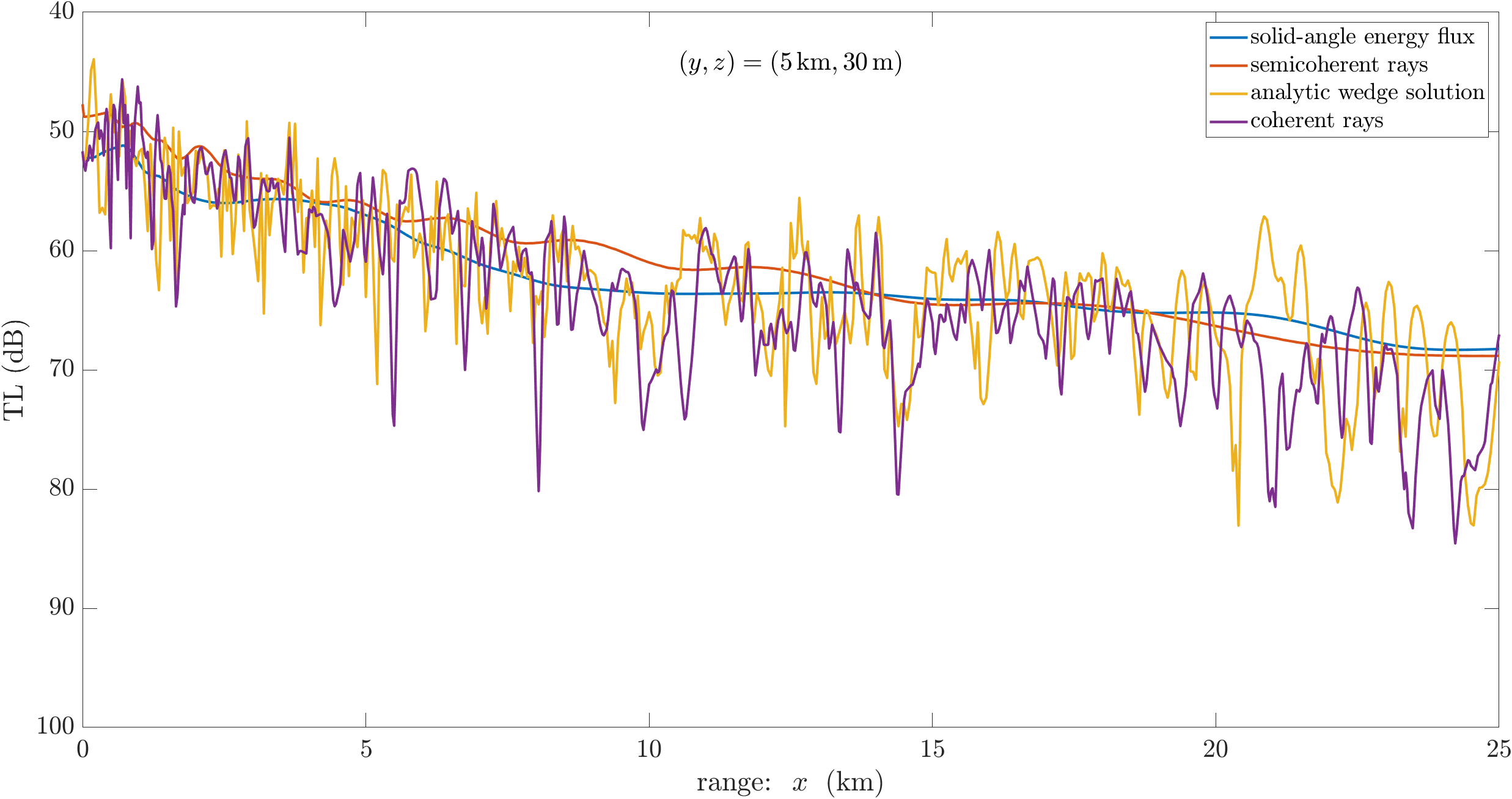}
  \caption{Comparison of TL for the ideal wedge I at $75\unit{Hz}$ frequency along a line array of receivers at $(y,z)=(5\unit{km},30\unit{m})$.}
  \label{fig:wej1f075z30y5000}
\end{figure}

Then at a higher frequency of $75\unit{Hz}$ in \cref{fig:wej1f075z30}, more wedge modes are present and still distinguishable, but overall the TL structure is more incoherent and in better agreement between all models.

\FloatBarrier

\subsection{Lossy Penetrable Wedge III} \label{ssec:wej3}

\begin{figure}[!htb]
  \includegraphics[width=\textwidth]{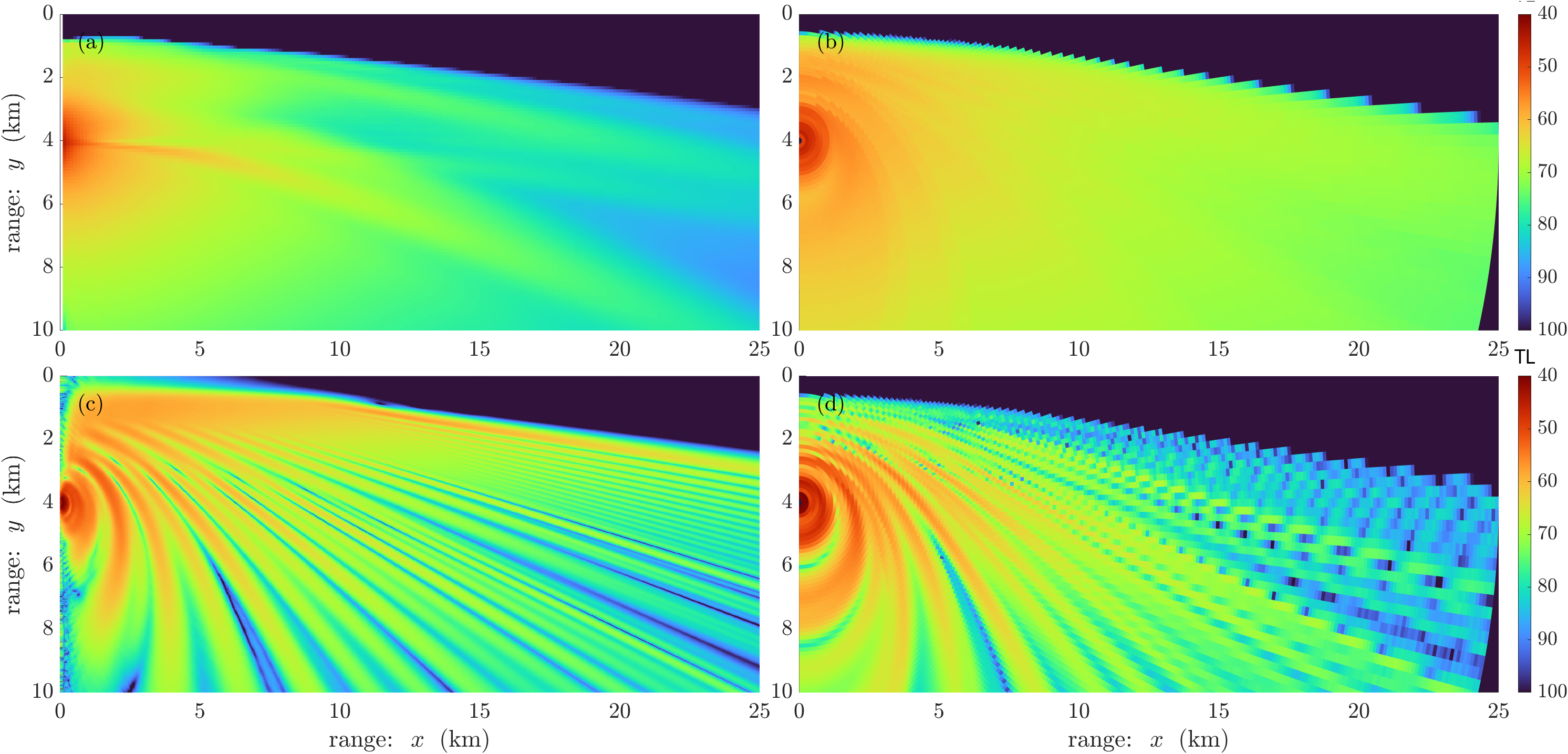}
  \caption{Side-by-side comparison of a 2D TL slice at $z=30\unit{m}$ depth for the lossy penetrable wedge III at $25\unit{Hz}$ frequency.  The models included are: (a) the 3D solid-angle energy flux model, (b) the semi-coherent 3D ray tracing model, (c) the 3D split-step Fourier wide-angle parabolic equation model, and (d) the coherent 3D ray tracing model.}
  \label{fig:wej3f025z30}
  
  \includegraphics[width=\textwidth]{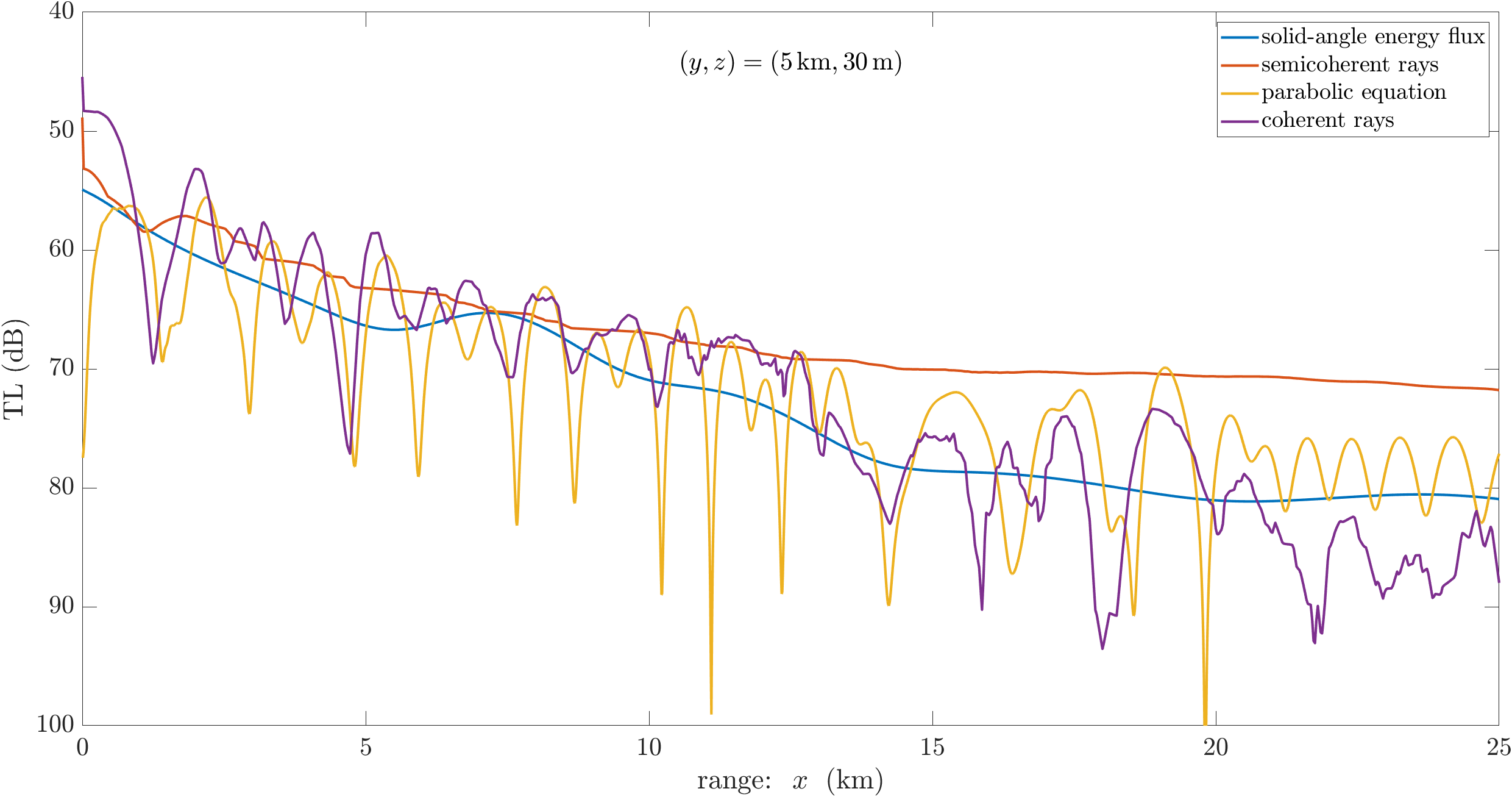}
  \caption{Comparison of TL for the lossy penetrable wedge III at $25\unit{Hz}$ frequency along a line array of receivers at $(y,z)=(5\unit{km},30\unit{m})$.}
  \label{fig:wej3f025z30y5000}
\end{figure}

\begin{figure}[!htb]
  \includegraphics[width=\textwidth]{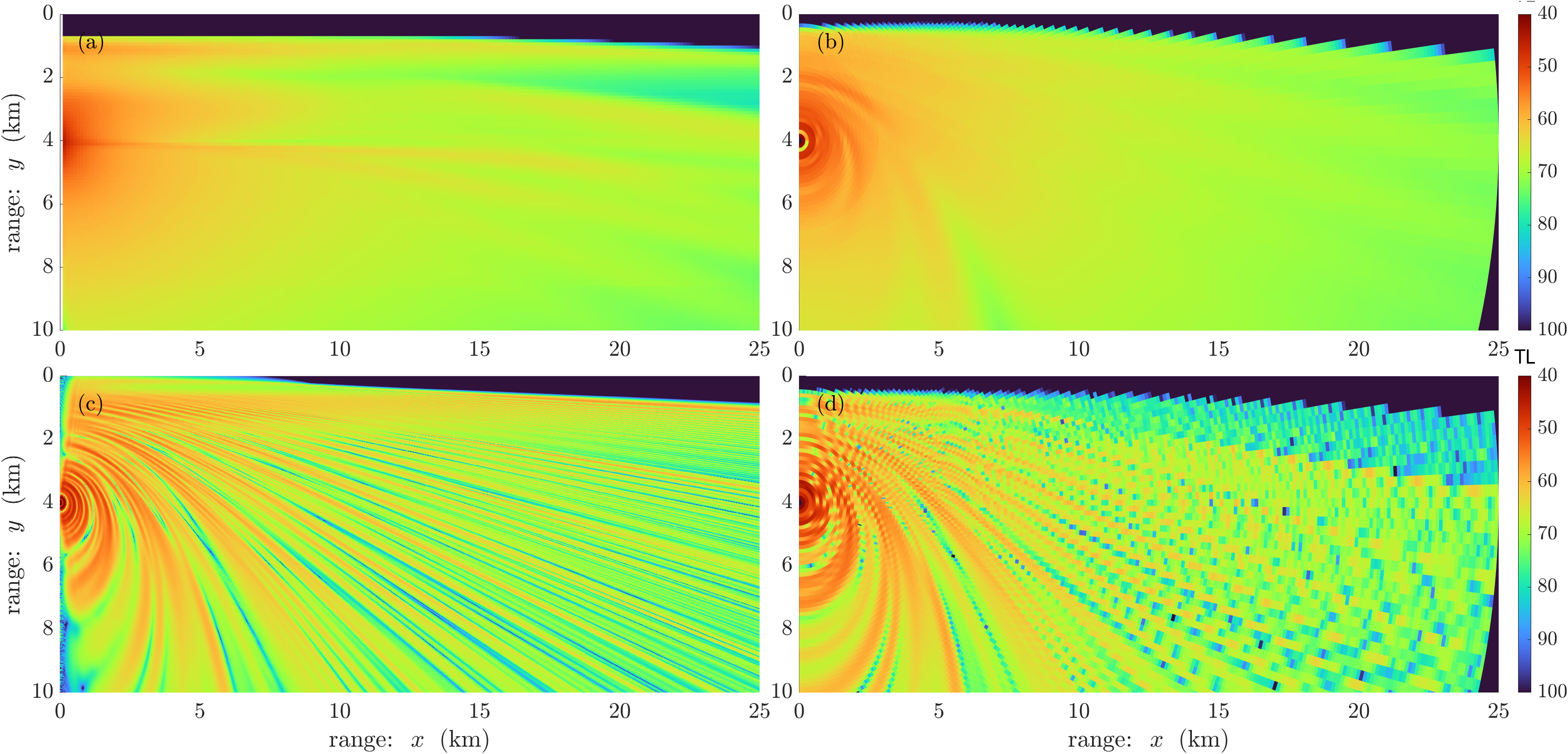}
  \caption{Side-by-side comparison of a 2D TL slice at $z=30\unit{m}$ depth for the lossy penetrable wedge III at $75\unit{Hz}$ frequency.  The models included are: (a) the 3D solid-angle energy flux model, (b) the semi-coherent 3D ray tracing model, (c) the 3D split-step Fourier wide-angle parabolic equation model, and (d) the coherent 3D ray tracing model.}
  \label{fig:wej3f075z30}

  \includegraphics[width=\textwidth]{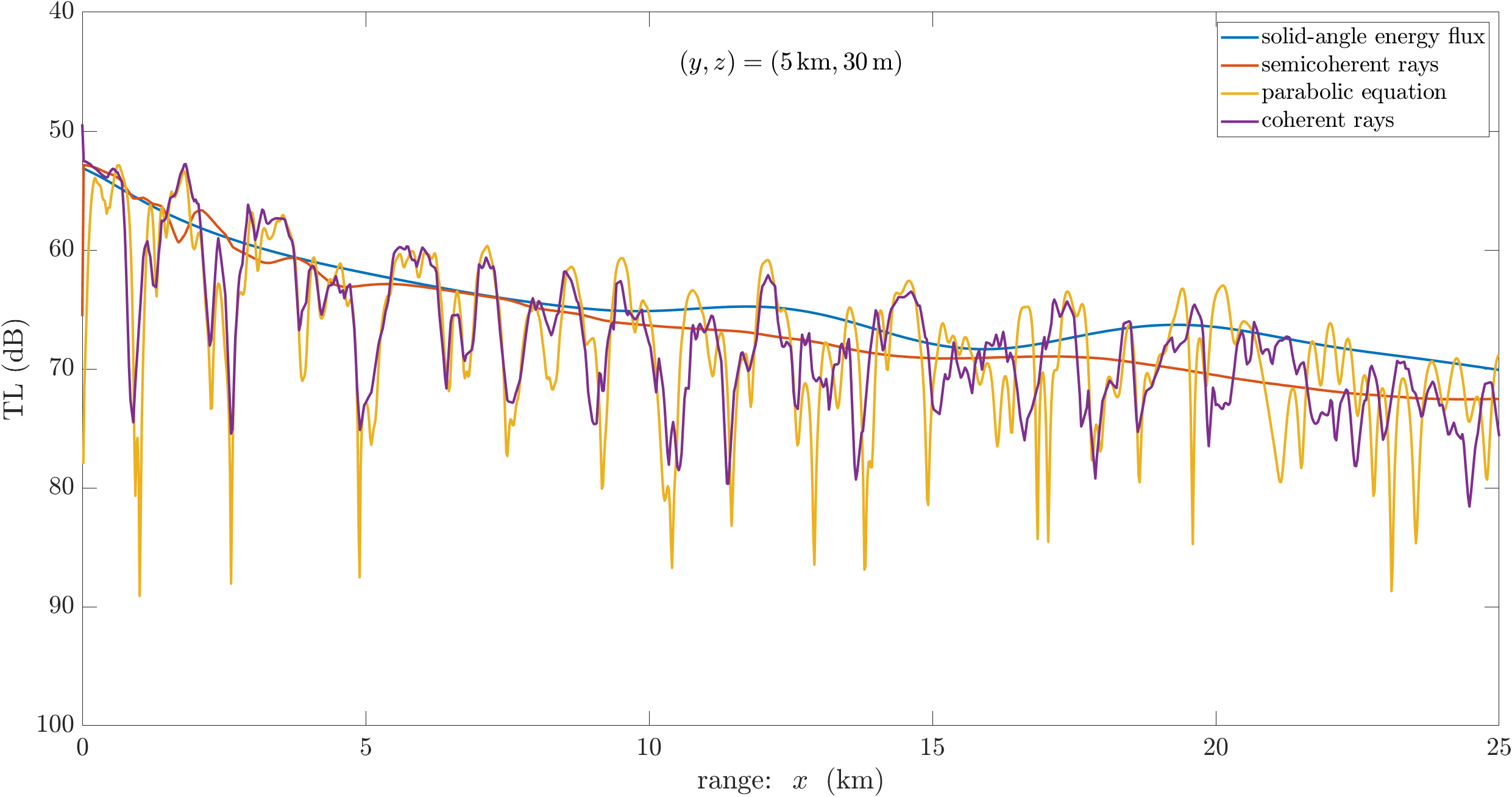}
  \caption{Comparison of TL for the lossy penetrable wedge III at $75\unit{Hz}$ frequency along a line array of receivers at $(y,z)=(5\unit{km},30\unit{m})$.}
  \label{fig:wej3f075z30y5000}
\end{figure}

\begin{figure}[!htb]
  \includegraphics[width=\textwidth]{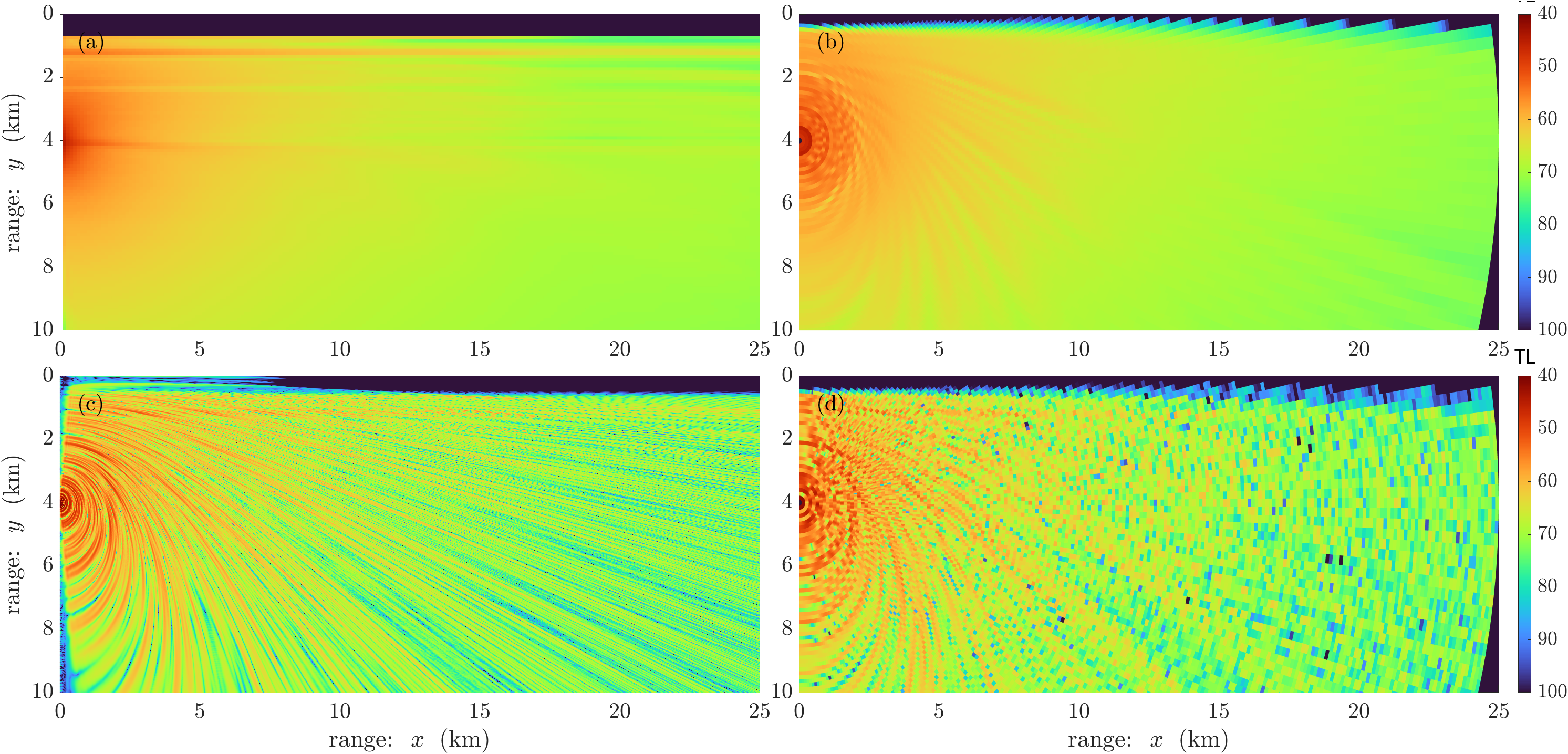}
  \caption{Side-by-side comparison of a 2D TL slice at $z=30\unit{m}$ depth for the lossy penetrable wedge III at $250\unit{Hz}$ frequency.  The models included are: (a) the 3D solid-angle energy flux model, (b) the semi-coherent 3D ray tracing model, (c) the 3D split-step Fourier wide-angle parabolic equation model, and (d) the coherent 3D ray tracing model.}
  \label{fig:wej3f250z30}
\end{figure}

\begin{figure}[!htb]
  \includegraphics[width=\textwidth]{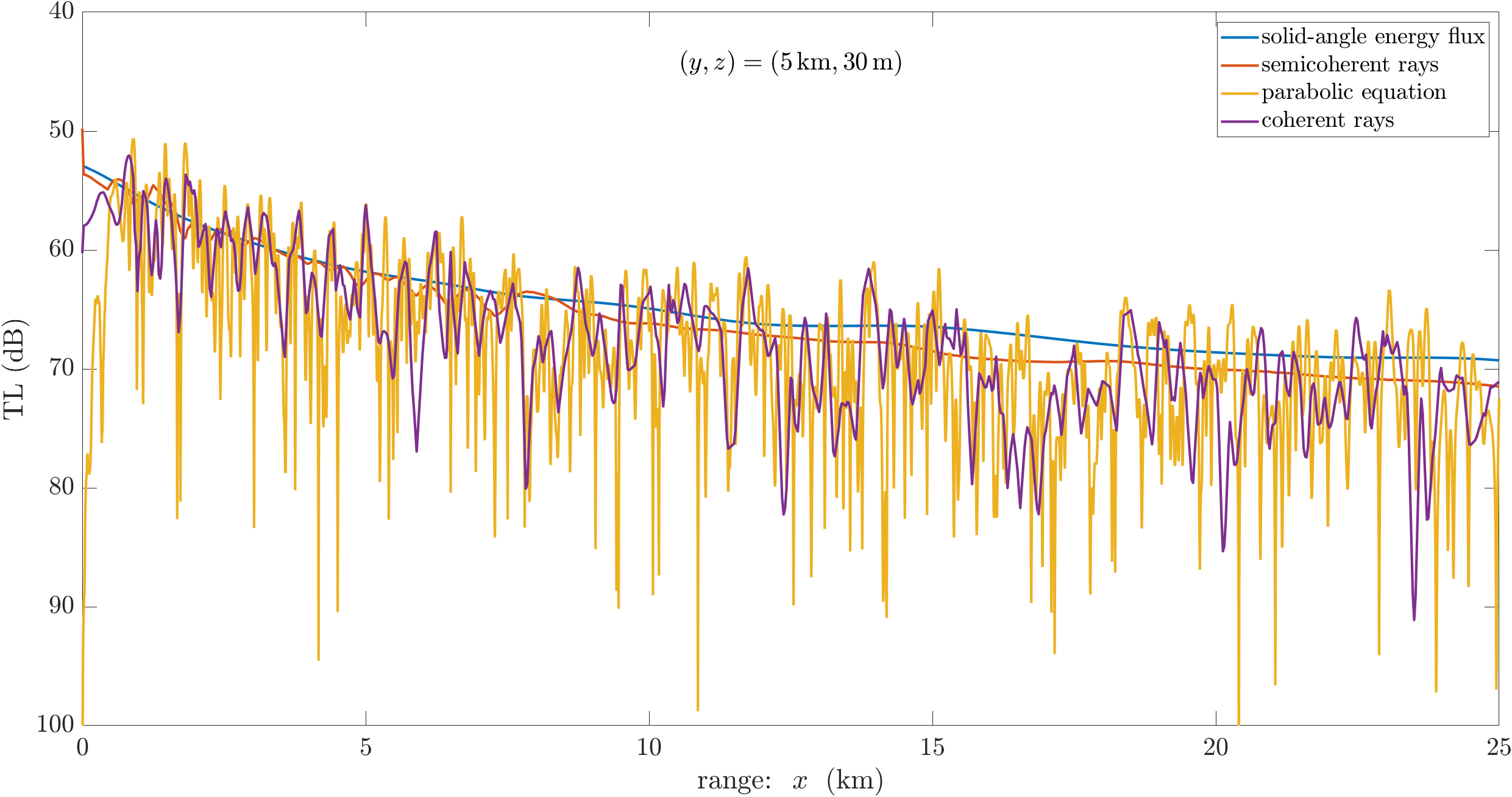}
  \caption{Comparison of TL for the lossy penetrable wedge III at $250\unit{Hz}$ frequency along a line array of receivers at $(y,z)=(5\unit{km},30\unit{m})$.}
  \label{fig:wej3f250z30y5000}
\end{figure}

For all side-by-side comparisons of 2D TL plots in this lossy penetrable wedge subsection, solution (a) in the top left corner is from the 3D solid-angle energy flux model with transverse $y$-convergence only, solution (b) in the top-right corner is the semicoherent ray-tracing solution calculated by Bellhop3D\supercite{porter2011bellhop}, solution (c) in the bottom-left corner is computed with a 3D wide-angle split-step-Fourier parabolic equation model developed by Ying-Tsong Lin\supercite{lin2013higher}, and solution (d) in the bottom-right corner is the coherent ray-tracing solution calculated by Bellhop3D\supercite{porter2011bellhop}.  Horizontal slices of the TL field are shown for 3 different frequencies: at $25\unit{Hz}$ in \cref{fig:wej3f025z30,fig:wej3f025z30y5000}, at $75\unit{Hz}$ in \cref{fig:wej3f075z30,fig:wej3f075z30y5000}, and at $250\unit{Hz}$ in \cref{fig:wej3f250z30,fig:wej3f250z30y5000}.

For the lossy penetrable wedge, the wedge modes are now less distinguishable as the rate of bottom loss begins to dominate over geometric spreading after the waves propagate to any appreciable horizontal range.  Again at the lower $25\unit{Hz}$ frequency, the semi-coherent ray tracing solution appears to be underestimating the incoherent TL levels at range.

The lossy penetrable wedge energy flux TL in \cref{fig:wej3f025z30,fig:wej3f075z30,fig:wej3f250z30} has a noticeable horizontal intensity band or caustic-like effect, and it is likely an artifact specific to the energy flux method's implementation.  The exact cause of this intensity band artifact is currently unknown, but it could be related to: (a) the splitting of the energy flux integration domain about $(\tht_0,\phi_0)=(0,0)$, and/or (b) the known phenomenon of horizontal intensity bands at the source position in energy flux solutions\supercite{weston1980wave,harrison2015efficient} which are much more distinguished in isovelocity environments.

Once again at higher frequencies, \cref{fig:wej3f250z30,fig:wej3f250z30y5000}, the models tend to converge toward an incoherent solution dominated by the average rate of bottom attenuation.

\FloatBarrier

\subsection{3D Adiabatic Rays Demo} \label{ssec:quadRays}

\begin{figure}[!htb]
  \includegraphics[width=\textwidth]{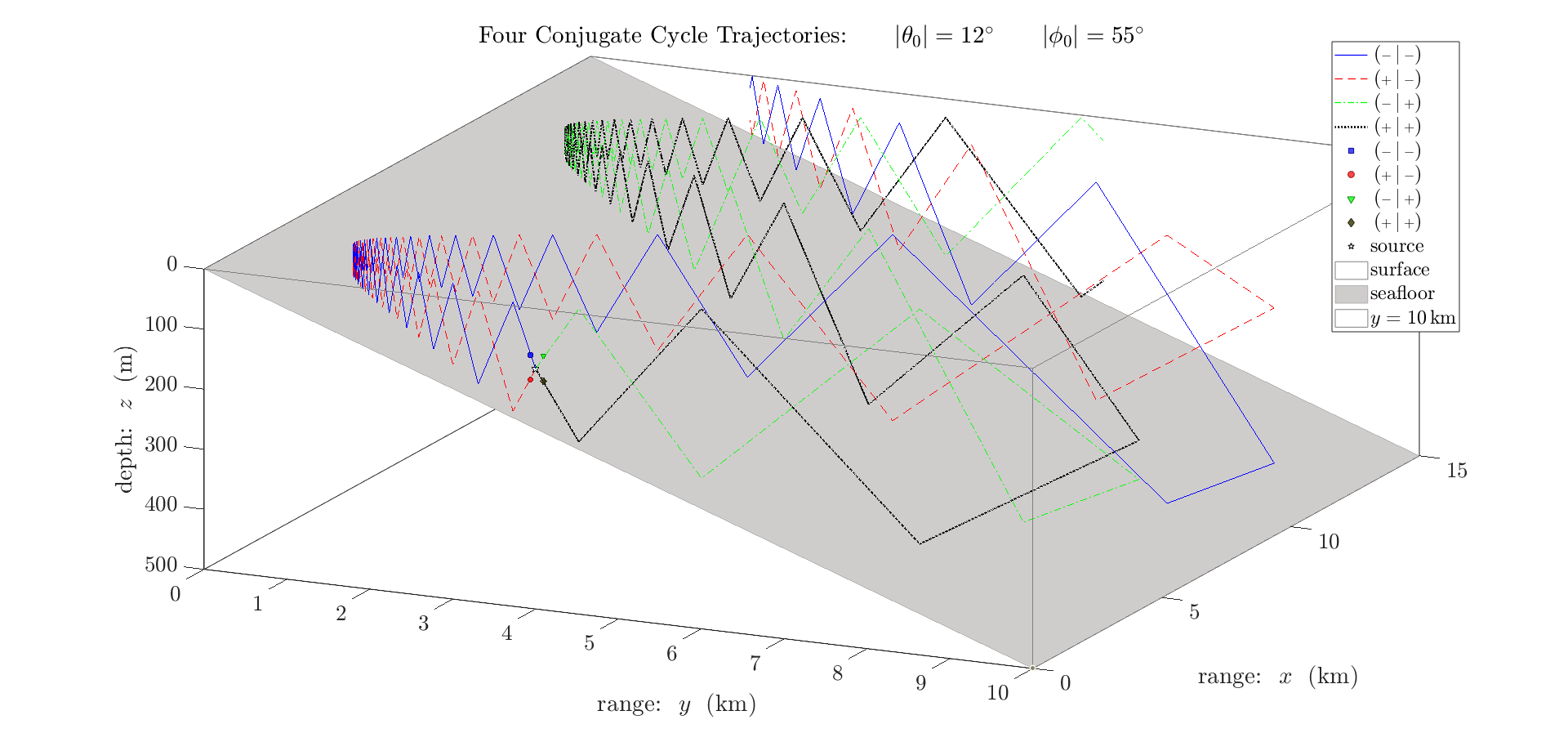}
  \caption{Four sign-conjugate 3D adiabatic cycle trajectories in the ideal wedge I at $25\unit{Hz}$ with a rigid wall at $y=10\unit{km}$ and initial propagation angle magnitudes $\abs{\(\theta_0,\phi_0\)}=\(12\deg,55\deg\)$.  The 3D cycle trajectories are distinguished by color, line style, and marker shape.  The orientations of the initial propagation angles are denoted as $\sgn\(\tht_0|\phi_0\)=\(\pm|\pm\)$.}
  \label{fig:wej1f025quadRays}
\end{figure}

One of the most interesting features of this 3D solid-angle energy flux model is the ability to simulate fully three-dimensional adiabatic cycle trajectories.  Even though the appropriate implementation of the vertical convergence factor is still currently being investigated, there is compelling evidence that the differential chains used to derive the vertical cycle tracking in \cref{eq:verticalCyclePosition,eq:verticalCycleMomentum} are correct.  This can be demonstrated with a test of the 3D cross-domain cycle tracking in an ideal wedge waveguide, and \cref{fig:wej1f025quadRays} shows four sign-conjugate cycle trajectories emanating from a source position located at $(y_0,z_0)=(4\unit{km},100\unit{m})$ in this environment.

These trajectories are not calculated by a marching method as is typically done in ray tracing models, but instead are interpolated from inverting mappings between partial cycle integrals and positions within the waveguide.  The 3D trajectories appear to bounce in straight line segments between the boundaries because of the isovelocity water-column, but they are actually curving very slightly both horizontally and vertically as the waveguide depth varies with $y$, a consequence of the adiabatic modes approximation.

The computational domain's transverse boundary at $y=10\unit{km}$ acts like a reflective boundary, because the model's derivation assumes vertical and transverse adiabatic cycles that are periodic in $x$ due to longitudinal $x$-range-independence.  The cycle trajectories could be explicitly turned off when they are incident upon the transverse boundaries, which is essentially how the transverse mode-stripping kernel in \cref{eq:transverseModeStrippingKernel} works to simulate geometric spreading when combined with the convergence factor.

On the apex side of the waveguide domain, all of the waves would be refracted at some minimum turning point in $y$ since all non-zero vertical WKB modenumbers have a non-zero minimum waveguide depth in which they can propagate that is determined by the vertical WKB phase integral. The specific shape of the horizontal trajectory and rate of horizontal refraction is governed by the adiabatic approximation which defines the effective transverse wavenumber profiles.

\begin{figure}[!htb]
  \includegraphics[width=\textwidth]{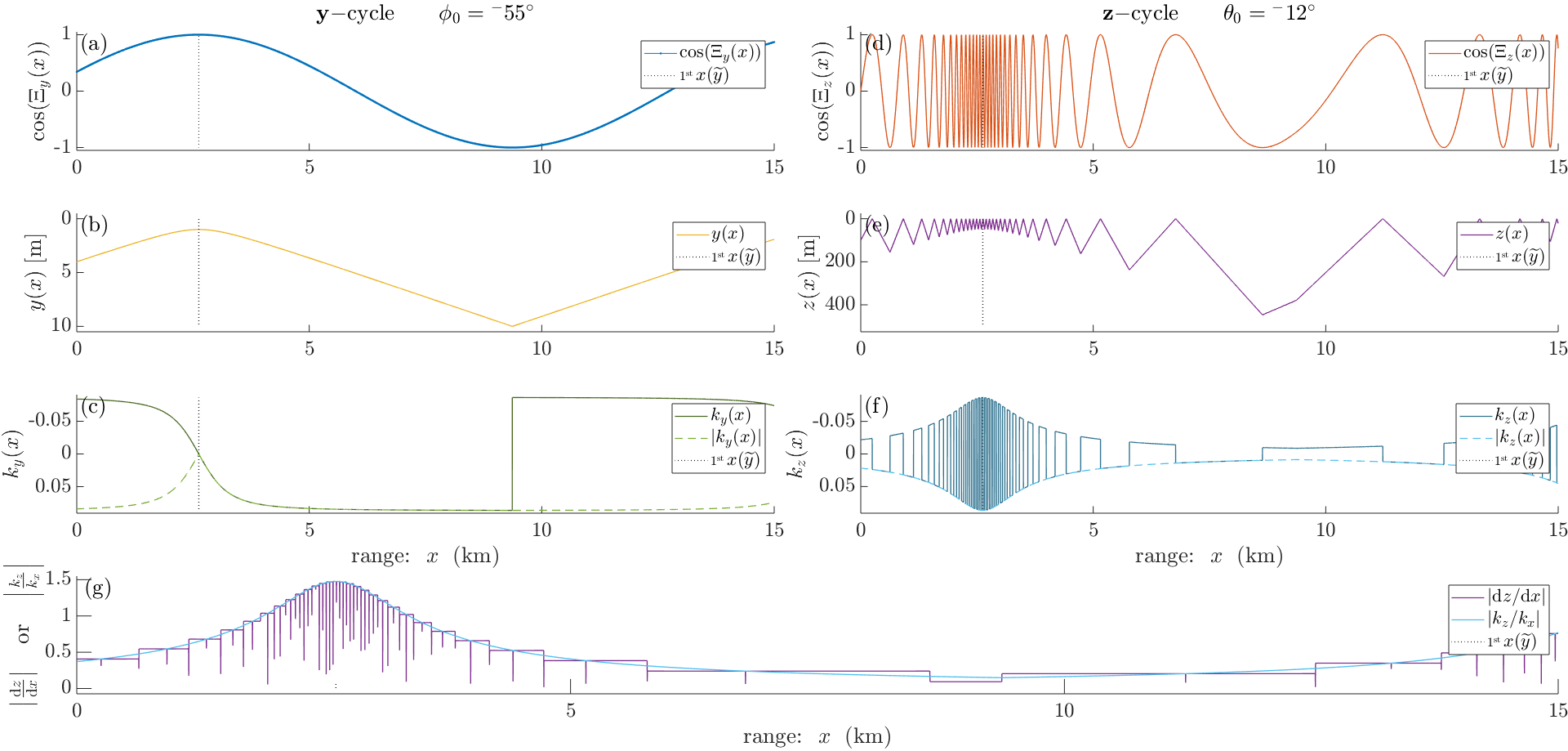}
  \caption{Transverse and vertical cycles as functions of $x$ for the initial propagation angle $\(\tht_0,\phi_0\)=\(-12\deg,-55\deg\)$ at $25\unit{Hz}$. Subfigures (a) and (d) show the cosines of the cycle phases, $\Xi_y(x)$ and $\Xi_z(x)$, for the transverse and vertical cycles respectively.  Subfigures (b) and (e) show the cycle positions, $y(x)$ and $z(x)$, for the transverse and vertical cycles respectively.  Subfigures (c) and (f) show the cycle momentums (wavenumber components), $k_y(x)$ and $k_z(x)$, for the transverse and vertical cycles respectively.  Subfigure (g) compares the divided difference $\abs{\dz/\dx}$ to the wavenumber component ratio $\abs{k_z/k_x}$.}
  \label{fig:wej1f025soloRayPhs}
\end{figure}

An important feature of these cross-domain cycle trajectories is how the rate and frequency of vertical cycling is modulated by the wave's position within its transverse cycle.  As a vertical mode propagates to shallower water, its vertical phase speed slows and the vertical cycle integral is evaluated over a shorter water-column, which is a parallel interpretation for a ray trajectory steepening and cycling faster between reflective boundaries that are closer together.

\Cref{fig:wej1f025soloRayPhs} shows detailed transverse and vertical cycle evolutions for one of the trajectories depicted in \cref{fig:wej1f025quadRays}.  The transverse cycle evolution is plotted in the left column, and the vertical cycle evolution is plotted in the right column.  Sinusoids of the cycle phases are plotted in the top row to highlight the periodic nature of the adiabatic cycle trajectories.  Cycle positions are plotted in the next row down, and cycle momentums or wavenumber components are plotted in the third row from the top.  And the bottom row shows a comparison of divided differences in vertical ($z$) and longitudinal ($x$) positions with the ratio of vertical and longitudinal wavenumber components, $k_z$ and $k_x$ respectively.  The first transverse turning point is denoted in all of the plots by a dotted vertical line, and the modulation of the vertical cycle by the transverse cycle is symmetric about this point.

\FloatBarrier

\section{DISCUSSION}
\label{sec:discussion}

\Cref{fig:wej1f025z30} demonstrates clearly that the model is accurately predicting the rate of horizontal refraction of the wedge modes.  In every slice of the TL in the plots above, it is evident that the energy flux solution predominantly follows the locally-averaged field levels of the reference solutions via analytic integral transform and parabolic equation methods.  It is curious that the semi-coherent ray tracing solutions seem to underestimate the background incoherent TL solutions at very low frequencies, but the energy flux solutions, the coherent ray solutions, and the higher frequency solutions do not have this discrepancy.  There are some computational artifacts present in the energy flux solutions that are likely systemic and inherent to the method's approximations, but they are expected since 2D energy flux models can exhibit similar artifacts such as intensity bands radiating and diverging from the source location.

If properly implemented with efficient algorithms in a low-level language as a production model, this 3D solid-angle energy flux model could be significantly more efficient at computing incoherent horizontally-refracted 3D acoustic fields in complex environments.  Since the energy flux field solution is built upon globally-evaluated adiabatic cycle integrals that are numerically sampled, interpolated, and inverted across the environment's computational domain, the cycle integrations depending only on the environmental parameters could potentially be analytically extracted from the field calculations and precomputed separately as lookup tables.  This may be done by isolating and factoring out the frequency dependence, which may require normalizing all parametric domains and compositing non-linear integration functions to account for the frequency and propagation angle dependence of the cycle integration limits.  However, it should be noted that the position of the integration limits do vary linearly locally with the vertexing sound speeds or wavenumbers if the profiles are themselves interpolated as piecewise linear.

To focus the development efforts on the primary cause of horizontal refraction in ocean environments, this version of the 3D solid-angle energy flux model that was implemented for this project was simplified in several ways: (a) the water-column is assumed to be vertically $z$-depth-independent and transversely $y$-range-independent, (b) the bottom boundary plane-wave reflection coefficient is assumed to be $y$-range-independent, (c) the entire waveguide environment is assumed to be $x$-range-independent, and (d) the model assumes that both vertical and transverse wavenumber profiles are convex.  Relaxing assumptions (a) and (b) should already be theoretically accounted for in the model's derivation, and are therefore straight-forward to implement with proper attention to detail.

It is likely possible to relax assumption (c) by implementing the same type of forward adiabatic range-dependence that is already used in 2D range-dependent energy flux models\supercite{weston1980acoustica,weston1980acousticb,harrison2015efficient}.  However, there may be hidden nuance and complications when the vertical and transverse cycles are allowed to adiabatically vary with the forward longitudinal $x$-range, and the algorithmic book-keeping would certainly become more complex.  In particular, careful consideration of the 3D cycle-tracking method and convergence factor is needed to ensure that the algorithms are physically and geometrically sensible.

With the $x$-range-dependence approach typically used in 2D energy flux models, there is still a limitation of no back-propagation along the forward $x$-direction since the adiabatic cycles are parameterized and evolved along this coordinate.  In order to extend the model to completely generalized three-dimensional propagation, another parameterizing variable for the energy flux model's evolution would have to be defined such as arclength, 3D cycle phase, or elapsed time.  The ray-mode analogy allows for intuitive interpretation of the derivation and solution in geometric terms, and this line of thought points to an important connection between the energy flux methods and dynamical Lagrangian systems, a connection first highlighted explicitly by Milder in 1969\supercite{milder1969ray}.

Assumption (d) has so far been universally true for all numerically implemented energy flux models due to the complications that can arise when constructing geometric wave solutions with bimodal potential wells. It may be possible to construct an energy flux model using the full WKB mode solutions expressed as complex exponentials, which would allow for mode functions spanning both the oscillatory and evanescent domains corresponding respectively to real and imaginary modal eigenvalues.  This could provide a way to properly capture and express wave-tunneling across potential barriers in non-convex wavenumber profiles, but it would likely require integrating complex propagation angles and the theoretical implications for this kind of approach are largely unexplored.

\section{SUMMARY}\label{sec:summary}

The primary objective of this project was to develop a 3D underwater acoustic propagation model that can capture the effects of horizontal refraction built exclusively on energy flux principles.  This objective was achieved and the model's primary capabilities were demonstrated, which includes resolving horizontally refractive acoustic fields and constructing 3D adiabatic ray cycles based on inverting mappings of cycle positions to partial cycle distances.  This model that was implemented is the first practical 3D energy flux model capable of calculating the averaged acoustic field in generalized 3D waveguides.  Though the model still has some limitations, it demonstrates the extensibility and generalizability of energy flux methods to more complex ocean acoustic problems.  The theoretical and analytical work performed also shows the explanatory power and physical insights that are possible when working with energy flux models, the ray-mode analogy, and the adiabatic approximation.

Several developments for energy flux methods were necessary to make this type of model possible, including: (a) geometric mode-stripping at transparent boundaries to simulate free propagation and geometric spreading, (b) unfolding the energy flux integral's angular domain by accounting for and distinguishing wave orientation in the cycle integrals and constructing asymmetric integration kernels, (c) tracking adiabatic cycles with accumulated cycle phase offsets by numerically inverting mappings between wave position and oriented partial cycle integrals, and (d) mapping the differential topology of the adiabatic double mode-sum in order to construct the appropriate differential chains and Jacobian transformations for the modal continuum.  These techniques make possible the construction of generalizable solid-angle integration kernels that account for specific 3D wave propagation phenomena that are physically and geometrically interpretable under the ray-mode analogy.

Efforts are still on-going to test and validate this 3D solid-angle energy flux model, and there is also additional development work needed to implement the full capability and generality of this model's derivation.  With continued development efforts, energy flux models could be a practical and efficient solution for computing the refractive structure of incoherent acoustic fields in realistic 3D undersea environments.


\section*{Acknowledgements}

%

This work was supported in part by the National Defense Science and Engineering Graduate (NDSEG) Fellowship Program, sponsored by the Air Force Office of Scientific Research (AFOSR), the Army Research Office (ARO), and the Office of Naval Research (ONR) under the Office of the Under Secretary for Research and Engineering, (OUSD R\&E).

%

This work was supported in part by the ARL Walker Fellowship Program, sponsored by the Applied Research Laboratory (ARL) at the Pennsylvania State University (PSU).

The authors also wish to acknowledge Sheri Martinelli and Dan Brown for their participation in technical discussions and general oversight during the course of this Ph.D. dissertation project.

\section*{Data Availability}

The 3D solid-angle energy flux model implementation that is used in this paper is licensed as open-source software under the MIT License and is hosted on Github: 
\url{https://github.com/marklanghirt/Tethys}\supercite{langhirt2025tethys}.

\printbibliography 

\end{document}

%% file: packages.tex
\usepackage[T1]{fontenc}
\usepackage[left=.5in, right=.5in, top=.5in, bottom=.5in]{geometry}

\usepackage{mathtools,amssymb}
\usepackage[bbgreekl]{mathbbol}
\usepackage{bm}  
\usepackage{mathrsfs} 
\usepackage{authblk}
\usepackage{setspace}
\usepackage{relsize}
\usepackage{nicefrac}
\usepackage{xparse} 
\usepackage{cancel} 
\usepackage{accents} 
\usepackage{outlines}
\usepackage{soul} 
\usepackage{placeins} 
\usepackage[backend=biber,bibstyle=numeric,citestyle=numeric-comp,sorting=none]{biblatex}
\usepackage{xcolor}
\usepackage{graphicx}
\usepackage{hyperref} 
\usepackage{cleveref} 

%% file: definitions.tex
\newcommand{\q}{\quad}
\newcommand{\qq}{\qquad}
\newcommand{\bvec}[1]{\boldsymbol{#1}}

\newcommand{\wtil}[1]{\widetilde{#1}}
\newcommand{\spacedline}[1]{\noalign{\vspace{#1pt}}\hline\noalign{\vspace{#1pt}}}

\renewcommand{\inf}{\infty}
\renewcommand{\min}{\bot}
\renewcommand{\max}{\top}
\newcommand{\lwr}{\smile}
\newcommand{\upr}{\frown}
\newcommand{\bigo}[1]{\mathscr{O}\(#1\)}

\renewcommand{\deg}{\ensuremath{{}^\circ}}
\newcommand{\pstv}{{}^+}
\newcommand{\ngtv}{{}^-}

\newcommand{\abs}[1]{\left|#1\right|}
\newcommand{\nfrac}[2]{\nicefrac{#1}{#2}}
\newcommand{\evalat}[2]{\left.#1\right|_{#2}}
\newcommand{\evalover}[3]{\left[#1\right]_{#2}^{#3}}
\newcommand{\lap}{\nabla^2}

\newcommand{\Real}[1]{\mathrm{Re}\left\{#1\right\}}
\DeclareMathOperator{\asin}{asin}
\DeclareMathOperator{\acos}{acos}
\DeclareMathOperator{\sgn}{sgn}
\DeclareMathOperator{\opmax}{max}
\DeclareMathOperator{\opmin}{min}

\NewDocumentCommand{\fd}{!O{}!O{}}{\,\mathrm{\del}_{#1}^{#2}{}}
\NewDocumentCommand{\fdd}{O{}mm!O{}}{\dfrac{\fd[][#1#4] #2}{\fd[][] #3^{#1#4}}}
\NewDocumentCommand{\D}{!O{}!O{}}{\,\mathrm{D}_{#1}^{#2}{}}
\NewDocumentCommand{\DD}{O{}mm!O{}}{\dfrac{\D[][#1#4] #2}{\D[][] #3^{#1#4}}}
\NewDocumentCommand{\nDD}{O{}mm!O{}}{\nicefrac{\D[][#1#4] #2}{\D[][] #3^{#1#4}}}
\NewDocumentCommand{\tDD}{O{}mm!O{}}{\tfrac{\D[][#1#4] #2}{\D[][] #3^{#1#4}}}

\NewDocumentCommand{\p}{!O{}!O{}}{\,\partial_{#1}^{#2}{}}
\NewDocumentCommand{\pp}{O{}mm!O{}}{\dfrac{\p[][#1#4] #2}{\p[][] #3^{#1#4}}}
\NewDocumentCommand{\npp}{O{}mm!O{}}{\nicefrac{\p[][#1#4] #2}{\p[][] #3^{#1#4}}}
\NewDocumentCommand{\tpp}{O{}mm!O{}}{\tfrac{\p[][#1#4] #2}{\p[][] #3^{#1#4}}}

\RenewDocumentCommand{\d}{!O{}!O{}}{\,\mathrm{d}_{#1}^{#2}{}}
\NewDocumentCommand{\dd}{O{}mm!O{}}{\dfrac{\d[][#1#4] #2}{\d[][] #3^{#1#4}}}
\NewDocumentCommand{\ndd}{O{}mm!O{}}{\nicefrac{\d[][#1#4] #2}{\d[][] #3^{#1#4}}}
\NewDocumentCommand{\tdd}{O{}mm!O{}}{\tfrac{\d[][#1#4] #2}{\d[][] #3^{#1#4}}}

\newcommand{\dk}{\d k}

\newcommand{\dx}{\d x}

\newcommand{\dy}{\d y}
\newcommand{\dz}{\d z}
\newcommand{\dtht}{\d\tht}

\newcommand{\dxi}{\d\xi}
\newcommand{\dphi}{\d\phi}
\newcommand{\dups}{\d\ups}
\newcommand{\dzet}{\d\zet}

\renewcommand{\(}{\left(}
\renewcommand{\)}{\right)}
\renewcommand{\[}{\left[}
\renewcommand{\]}{\right]}

\newcommand{\pmat}[1]{\begin{pmatrix}#1\end{pmatrix}}
\newcommand{\bmat}[1]{\begin{bmatrix}#1\end{bmatrix}}

\newcommand{\raro}{\rightarrow}
\newcommand{\Raro}{\Rightarrow}

\newcommand{\unit}[1]{\,\ensuremath{\mathsmaller{\mathrm{#1}}}}
\newcommand{\tand}{\q\text{and}\q}

\newcommand{\Let}{\text{Let:}\q}

\newcommand{\tfor}{\q\text{for}\q}
\newcommand{\twhere}{\q\text{where}\q}

\newcommand{\comma}{,\q}
\newcommand{\command}{,\q\text{and}\q}

\newcommand{\FrR}{\mathfrak{R}}

\newcommand{\SL}{\mathscr{L}}

\newcommand{\TL}{\mathrm{TL}}
\newcommand{\CC}{\mathcal{C}}
\newcommand{\CD}{\mathcal{D}}
\newcommand{\CM}{\mathcal{M}}
\newcommand{\CP}{\mathcal{P}}
\newcommand{\CR}{\mathcal{R}}

\newcommand{\CZ}{\mathcal{Z}}

\newcommand{\alf}{\alpha}
\newcommand{\bet}{\beta}
\newcommand{\del}{\delta}

\newcommand{\veps}{\varepsilon}
\newcommand{\zet}{\zeta}
\newcommand{\tht}{\theta}
\newcommand{\kap}{\kappa}

\newcommand{\lam}{\lambda}
\newcommand{\Lam}{\Lambda}
\newcommand{\sig}{\sigma}
\newcommand{\vsig}{\varsigma}
\newcommand{\ups}{\upsilon}
\newcommand{\Omg}{\Omega}

\newcommand{\ID}{\mathbb{D}}
\newcommand{\IF}{\mathbb{F}}
\newcommand{\IL}{\mathbb{L}}
\renewcommand{\Im}{\mathbb{m}}
\newcommand{\IM}{\mathbb{M}}
\newcommand{\In}{\mathbb{n}}
\newcommand{\IN}{\mathbb{N}}
\newcommand{\IR}{\mathbb{R}}

\newcommand{\ILam}{\mathbb{\Lam}}
\newcommand{\IPhi}{\mathbb{\Phi}}
\newcommand{\IChi}{\mathbb{X}}

\newcommand{\half}{\tfrac{1}{2}}